\documentclass[11pt,preprint]{elsarticle}

\usepackage{fancyhdr}
\usepackage{fullpage}
\usepackage{amsmath}
\usepackage{amsbsy}
\usepackage{amssymb}
\usepackage{amscd}
\usepackage{amsfonts}
\usepackage{supertabular}
\usepackage{graphics}
\usepackage{verbatim}
\usepackage{epsfig}
\usepackage{xspace}
\usepackage{euscript}
\usepackage{alltt}
\usepackage{boxedminipage}
\usepackage{float}
\usepackage[colorlinks]{hyperref}
\usepackage{color}
\usepackage[all]{xy}
\usepackage{t1enc}
\usepackage{times,exscale}
\usepackage{graphicx,calc}
\usepackage{subfig}
\usepackage[ruled,vlined]{algorithm2e}
\usepackage{epstopdf}
\usepackage{multirow}
\usepackage{pseudocode}
\usepackage{braket}
\usepackage{esint}

\def\bR{{\mathbf{R}}}
\def\bx{{\mathbf{x}}}
\def\bg{{\mathbf{g}}}
\def\R{{\mathbb{R}}}
\def\C{{\mathbb{C}}}
\def\bk{{\mathbf{k}}}
\def\bkb{{\mathbf{k_b}}}

\journal{arXiv}

\begin{document}

\begin{frontmatter}

\title{SPARC: Accurate and efficient finite-difference formulation and parallel implementation of Density Functional Theory: Extended systems}
\author[gatech]{Swarnava Ghosh}
\author[gatech]{Phanish Suryanarayana\corref{cor}}
\address[gatech]{College of Engineering, Georgia Institute of Technology, GA 30332, USA}
\cortext[cor]{Corresponding Author (\it phanish.suryanarayana@ce.gatech.edu) }

\begin{abstract}
As the second component of SPARC (Simulation Package for Ab-initio Real-space Calculations), we present an accurate and efficient finite-difference formulation and parallel implementation of Density Functional Theory (DFT) for extended systems. Specifically, employing a local formulation of the electrostatics, the Chebyshev polynomial filtered self-consistent field iteration, and a reformulation of the non-local force component, we develop a finite-difference framework wherein both the energy and atomic forces can be efficiently calculated to within desired accuracies in DFT. We demonstrate using a wide variety of materials systems that SPARC achieves high convergence rates in energy and forces with respect to spatial discretization to reference plane-wave result; exponential convergence in energies and forces with respect to vacuum size for slabs and wires; energies and forces that are consistent and display negligible `egg-box' effect; accurate properties of crystals, slabs, and wires; and negligible drift in molecular dynamics simulations. We also demonstrate that the weak and strong scaling behavior of SPARC is similar to well-established and optimized plane-wave implementations for systems consisting up to thousands of electrons, but with a significantly reduced prefactor. Overall, SPARC represents an attractive alternative to plane-wave codes for performing DFT simulations of extended systems.  
\end{abstract}


\end{frontmatter}

\section{Introduction}
Density Functional Theory (DFT) developed by Hohenberg, Kohn, and Sham \cite{Hohenberg,Kohn1965} is a popular ab-initio method for understanding as well as predicting a wide range of materials properties \cite{jones1989density,ziegler1991approximate,kohn1996density,jones2015density}. However, the solution of the DFT problem still remains a formidable task, which severely restricts the size of systems that can be studied. The plane-wave discretization has been a popular choice \cite{VASP,CASTEP,ABINIT,Espresso,DFT++,gygi2008architecture} since it forms a complete and orthonormal set, provides spectral convergence with increasing basis size, enables the efficient evaluation of convolutions through the Fast Fourier Transform (FFT) \cite{Cooley1965}, and is amenable to efficient and effective preconditioning \cite{payne1992iterative,hutter1994electronic}. However, the need for periodicity makes the plane-wave basis unsuitable for the study of non-periodic and localized systems such as clusters, surfaces and wires \cite{natan2008real,Phanish2012}. Additionally, its non-local nature makes the efficient use of modern large-scale computer architectures particularly challenging \cite{bottin2008large,tuckerman2000exploiting}, and the development of methods that scale linearly with respect to the number of atoms impractical \cite{Goedecker,Bowler2012}. 

The aforementioned limitations of the plane-wave basis have motivated the development of various real-space DFT implementations, including finite-differences \cite{chelikowsky1994finite,OCTOPUS,briggs1996real,fattebert1999finite,shimojo2001linear,Iwata2010}, finite-elements \cite{Sterne,white1989finite,tsuchida1995electronic,Phanish2010,motamarri2012higher,fang2012kohn}, wavelets \cite{arias1999multiresolution,cho1993wavelets,genovese2008daubechies}, periodic sinc functions \cite{ONETEP}, basis splines (B-splines) \cite{CONQUEST}, non-uniform rational B-splines (NURBS) \cite{masud2012b}, and maximum entropy basis functions \cite{Phanish2011}. However, despite the success of real-space approaches, plane-wave implementations still remain the method of choice for practical DFT computations. This is mainly due to their superior performance in achieving chemical accuracies on the modest computational resources that are commonly available to researchers \cite{sparc2016cluster}. 

The finite-difference discretization is an attractive option for performing real-space DFT simulations because it generates a standard eigenvalue problem with relatively small spectral width, an attribute that is critical to the eigensolver performance, particularly in the absence of effective real-space preconditioners. In addition, high-order approximations can be easily incorporated, a feature that is essential for efficient electronic structure calculations. However, the lack of an underlying variational structure due to the absence of a basis can result in non-monotonic convergence of the energies and atomic forces. In addition, the reduced accuracy of spatial integrations due to the use of a lower order integration scheme can lead to a pronounced `egg-box' effect \cite{ono2010real,BobSchChe15}---phenomenon arising due to the breaking of the translational symmetry---which can significantly impact the accuracy of structural relaxations and molecular dynamics simulations \cite{andrade2015real,li2016atomicbasis,artacho2009}. 

In recent work \cite{sparc2016cluster}, we have developed an accurate and efficient finite-difference formulation and parallel implementation of DFT for isolated systems, which represents the first component of SPARC (Simulation Package for Ab-initio Real-space Calculations). The solution methodology in SPARC includes a local formulation of the electrostatics, the Chebyshev polynomial filtered self-consistent field iteration, and a reformulation of the non-local component of the force. The electrostatic formulation, atomic force calculation, and overall parallel implementation distinguishes SPARC from existing finite-difference DFT packages like PARSEC \cite{zhou2006parallel} and OCTOPUS \cite{OCTOPUS}. Notably, for isolated systems, we have found that SPARC significantly outperforms plane-wave codes like ABINIT \cite{ABINIT} as well as finite-difference codes like PARSEC and OCTOPUS for achieving the accuracy desired in electronic structure calculations \cite{sparc2016cluster}.  

In this work, we extend the capabilities of SPARC to enable the study of the static and dynamic properties of extended systems like crystals, surfaces, and wires.\footnote{Extended systems are infinite in one or more directions. Therefore, the reduction of the calculations to a finite unit-cell require the electrostatics to be carefully reformulated, mainly due to their long-ranged nature. In addition, Bloch-periodic boundary conditions on the orbitals and Brillouin zone integration need to be incorporated, which introduces significant additional complexity into the theoretical, numerical, and computational aspects of DFT formulations and implementations. Also, since the underlying physics of isolated clusters are inherently different from extended systems, a thorough investigation into the accuracy and efficiency of the developed DFT approaches/code for such systems is warranted. This provides the motivation for the current work.} Through selected examples, we demonstrate that SPARC obtains high rates of convergence in the energy and forces to reference plane-wave results on refining the discretization; exponential convergence in energies and forces with respect to vacuum size for slabs and wires; forces that are consistent with the energy, both being free from any noticeable `egg-box' effect; accurate ground-state properties; and negligible drift in molecular dynamics simulations. Moreover, we show that SPARC displays weak and strong scaling behavior that is similar to well-established plane-wave codes, but with a significantly smaller prefactor.

The remainder of this paper is organized as follows. In Section \ref{Section:KSDFT}, we provide the mathematical details of DFT for extended systems. In Section \ref{Sec:FormulationImplementation}, we describe its finite-difference formulation and parallel implementation in SPARC, whose accuracy and efficiency is verified through selected examples in Section \ref{Sec:Examples}. Finally, we provide concluding remarks in Section \ref{Sec:Conclusions}. 

\section{Density Functional Theory (DFT)} \label{Section:KSDFT}
Consider a unit cell $\Omega$ with $N$ atoms and a total of $N_e$ valence electrons. Let the nuclei be positioned at $\bR = \{\bR_1, \bR_2, \ldots, \bR_N \}$ and possess valence charges $\{Z_1, Z_2, \ldots, Z_N\}$, respectively. Neglecting spin, the system's free energy in Density Functional Theory (DFT) \cite{Hohenberg,Kohn1965} can be written as
\begin{equation} \label{Eqn:Energy}
\mathcal{F} (\Psi, \bg, \bR) =  T_s(\Psi,\bg) + E_{xc}(\rho) +  K(\Psi,\bg,\bR) + E_{el}(\rho,\bR) - TS(\bg) \,,
\end{equation}
where $\Psi = \{ \psi_1, \psi_2, \ldots, \psi_{N_s}\}$ is the collection of orbitals with occupations $\bg = \{g_1, g_2, \ldots, g_{N_s} \}$, $\rho$ is the electron density, and $T$ is the electronic temperature. The electron density itself can be expressed in terms of the orbitals and their occupations as
\begin{equation}
\rho(\bx) = 2 \sum_{n=1}^{N_s} \fint_{BZ} g_n(\bk) | \psi_n(\bx,\bk) |^2 \, \mathrm{d\bk} \,,
\end{equation}
where $\bk$ denotes the Bloch wavevector, and $\fint_{BZ}$ signifies the volume average over the Brillouin zone. In Eqn. \ref{Eqn:Energy}, the first term is the electronic kinetic energy, the second term is the exchange-correlation energy, the third term is the non-local pseudopotential energy, the fourth term is the total electrostatic energy, and the final term is the entropic contribution arising from the partial occupations of the orbitals. 

\paragraph{Electronic kinetic energy} In Kohn-Sham DFT, the kinetic energy of the non-interacting electrons takes the form
\begin{equation}
T_s(\Psi,\bg) = - \sum_{n=1}^{N_s} \fint_{BZ} \int_{\Omega} g_n(\bk) \psi_n^*(\bx,\bk) \nabla^2 \psi_n(\bx,\bk) \, \mathrm{d \bx} \, \mathrm{d\bk}  \,.
\end{equation}
where the superscript $*$ denotes the complex conjugate. 

\paragraph{Exchange-correlation energy} The exact form of the exchange-correlation energy is currently unknown. Therefore, a number of approximations have been developed, including the widely used Local Density Approximation (LDA) \cite{Kohn1965}: 
\begin{equation}
E_{xc} (\rho) = \int_{\Omega} \varepsilon_{xc} (\rho(\bx)) \rho(\bx) \, \mathrm{d \bx} \,,
\end{equation}
where $\varepsilon_{xc} (\rho) = \varepsilon_x (\rho) + \varepsilon_c (\rho)$ is the sum of the exchange and correlation per particle of a uniform electron gas. 

\paragraph{Non-local pseudopotential energy} The non-local pseudopotential energy within the Kleinman-Bylander \cite{kleinman1982efficacious} representation takes the form
\begin{equation}
K(\Psi,\bg,\bR) = 2 \sum_{n=1}^{N_s} \fint_{BZ} g_n(\bk) \sum_{J} \sum_{lm} \gamma_{Jl} \left| \sum_{J'} \int_{\Omega} \chi_{J'lm}^*(\bx,\bR_{J'}) e^{i \bk .(\bR_J-\bR_{J'})} \psi_n(\bx,\bk) \, \mathrm{d \bx} \right|^2 \, \mathrm{d\bk} \,,
\end{equation}
where the summation index $J$ runs over all atoms in $\Omega$, and the summation index $J'$ runs over the $J^{th}$ atom and its periodic images. In addition, the coefficients $\gamma_{Jl}$ and projection functions $\chi_{Jlm}$ can be expressed as
\begin{equation}
\gamma_{Jl} = \left( \int_{\R^3} \chi_{Jlm}^*(\bx,\bR_J) u_{Jlm}(\bx,\bR_J)  \, \mathrm{d\bx} \right)^{-1} \,, \,\, \chi_{J'lm}(\bx,\bR_{J'}) = u_{J'lm}(\bx,\bR_{J'}) \left(V_{J'l}(\bx,\bR_{J'})-V_{J'}(\bx,\bR_{J'}) \right) \,.
\end{equation}
Above, $u_{J'lm}$ are the isolated atom pseudowavefunctions, $V_{J'l}$ are the angular momentum dependent pseudopotentials, and $V_{J'}$ are the local components of the pseudopotentials, with $l$ and $m$ signifying the azimuthal and magnetic quantum numbers, respectively. 

\paragraph{Electrostatic energy} The total electrostatic energy is the sum of three components:  
\begin{equation}
E_{el}(\rho,\bR) = \frac{1}{2} \int_{\R^3} \int_{\Omega} \frac{\rho(\bx)\rho(\bx')}{|\bx - \bx'|} \,\mathrm{d\bx} \, \mathrm{d\bx'} + \sum_{I} \int_{\Omega} \rho(\bx) V_I(\bx,\bR_I)  \, \mathrm{d\bx} + \frac{1}{2} \sum_{I} \sum_{J \neq I} \frac{Z_{I} Z_{J}}{|\bR_{I}-\bR_{J}|}  \label{Eqn:ElectrostaticEnergy} \,,
\end{equation}
where the first term---referred to as the Hartree energy---is the classical interaction energy of the electron density, the second term is the interaction energy between the electron density and the nuclei, and the third term is the repulsion energy between the nuclei. The summation index $I$ runs over all atoms in $\R^3$, i.e., all the atoms in $\Omega$ and their periodic images.  

\paragraph{Electronic entropy} The electronic entropy originating from the partial orbital occupations:
\begin{equation} 
S(\bg) = - 2 k_B \sum_{n=1}^{N_{s}} \fint_{BZ} \left( g_n(\bk) \log g_n(\bk) + (1-g_n(\bk)) \log (1-g_n(\bk)) \right) \, \mathrm{d\bk} \,,
\end{equation}
where $k_B$ is the Boltzmann constant. 

\paragraph{Ground state/Molecular dynamics} The overall ground state in DFT is governed by the variational problem 
\begin{equation} \label{Eqn:GroundStateSplit}
\mathcal{F}_{0}  =   \inf_{\bR} \hat{\mathcal{F}}(\bR)  \,,
\end{equation} 
where
\begin{equation} \label{Eqn:GroundStateElectronic} 
\hat{\mathcal{F}}(\bR) = \inf_{ \Psi ,\bg} \mathcal{F}(\Psi,\bg,\bR) \, \quad s.t. \quad \int_{\Omega} \psi_i^*(\bx,\bk) \psi_j(\bx,\bk) \, \mathrm{d\bx} = \delta_{ij} \,, \quad 2 \sum_{n=1}^{N_s} \fint_{BZ} g_n(\bk) \, \mathrm{d\bk} = N_e  \,.
\end{equation}
In this staggered approach, the electronic ground-state as described by the above equation and the corresponding Hellmann-Feynman forces need to be computed for every configuration of the nuclei encountered during geometry optimization/molecular dynamics. 


\section{Formulation and implementation} \label{Sec:FormulationImplementation}
In this section, we describe the real-space formulation and parallel finite-difference implementation of Density Functional Theory (DFT) for extended systems. This represents the second component of SPARC (Simulation Package for Ab-initio Real-space Calculations) \cite{sparc2016cluster}, a first principles code currently under development for efficient large-scale electronic structure calculations.  

\paragraph{Electrostatic reformulation} The non-locality of the electrostatic energy in Eqn. \ref{Eqn:ElectrostaticEnergy} makes its direct real-space evaluation scale as $\mathcal{O}(N^2)$ with respect to the number of atoms. Furthermore, the individual components diverge in extended systems. To overcome this, we adopt a local formulation of the electrostatics \cite{Pask2005,ghosh2014higher}:
\begin{equation} \label{Eqn:ElecEnergyReformulation}
E_{el}(\rho,\bR) = \sup_{\phi}  \bigg \{ - \frac{1}{8 \pi} \int_{\Omega} |\nabla \phi(\bx,\bR)|^2 \, \mathrm{d\bx} + \int_{\Omega}(\rho(\bx)+ b(\bx,\bR)) \phi(\bx,\bR) \, \mathrm{d\bx} \bigg \} - E_{self}(\bR)  + E_c(\bR)  \,, 
\end{equation}
where $\phi$ denotes the electrostatic potential. In addition, $b$ represents the total pseudocharge density of the nuclei: 
\begin{eqnarray} 
b(\bx,\bR) = \sum_{I} b_I(\bx,\bR_I) \,, & & b_I(\bx,\bR_I) = - \frac{1}{4 \pi} \nabla^2 V_I(\bx,\bR_I) \,, \label{Eqn:PseudochargeDefinition} \\
\int_{\Omega} b(\bx,\bR) \, \mathrm{d\bx} = -N_e \,, & &  \int_{\R^3} b_I(\bx,\bR_I) \, \mathrm{d\bx} = Z_I  \,, \label{Eqn:IntegralPseudocharges}
\end{eqnarray}
where the summation index $I$ runs over all atoms in $\R^3$, and $b_I$ is the pseudocharge density of the $I^{th}$ nucleus that generates the potential $V_I$. The second to last term in Eqn. \ref{Eqn:ElecEnergyReformulation} represents the self energy associated with the pseudocharge densities:
\begin{equation} \label{Eqn:SelfEnergy}
E_{self}(\bR) = \frac{1}{2}\sum_{I} \int_{\Omega} b_I(\bx,\bR_I) V_I(\bx,\bR_I) \, \mathrm{d\bx} \,. 
\end{equation}
The last term $E_c$, whose explicit expression can be found in Appendix \ref{Appendix:Correct:RepulsiveEnergy}, corrects for the error in the repulsive energy when the pseudocharge densities overlap.

\paragraph{Electronic ground-state} The electronic ground-state for a given position of nuclei is governed by the constrained minimization problem in Eqn. \ref{Eqn:GroundStateElectronic}. On taking the first variation and utilizing Bloch's theorem \cite{bloch1929quantenmechanik}:
\begin{equation}
\psi_n(\bx,\bk) = e^{i \bk . \bx} u_n(\bx,\bk) \,,
\end{equation}
we arrive at
\begin{eqnarray} 
& & \left( \mathcal{H}\equiv -\frac{1}{2} \nabla^2 - i\bk. \nabla + \frac{1}{2} |\bk|^2 + V_{xc} + \phi + e^{-i \bk . \bx} V_{nl} e^{i \bk . \bx} \right)  u_n  = \lambda_n u_n \,, \quad n=1,2, \ldots, N_s \,, \nonumber \\
& &  g_n(\bk) = \left( 1 + \exp\left( \frac{\lambda_n(\bk) - \lambda_f}{k_B T} \right) \right)^{-1} \,, \quad \text{where} \,\, \lambda_f \,\,\, \text{is} \,\,\, s.t. \,\,\, 2 \sum_{n=1}^{N_s} \fint_{BZ} g_n(\bk) \, \mathrm{d \bk} = N_e \,, \label{Eqn:EL} \\
& & \rho(\bx) = 2 \sum_{n=1}^{N_s} \fint_{BZ} g_n(\bk) | u_n(\bx,\bk) |^2 \, \mathrm{d\bk} \,, \quad -\frac{1}{4 \pi} \nabla^2 \phi(\bx,\bR) = \rho(\bx) + b(\bx,\bR) \,,  \nonumber 
\end{eqnarray}
where $i=\sqrt{-1}$, $u$ is a function that is periodic on the unit cell in the directions of periodicity, $\mathcal{H}$ is the Hamiltonian operator, $V_{xc}= \delta E_{xc}/\delta \rho$ is the exchange-correlation potential, $\lambda_f$ is the Fermi energy, and $V_{nl}$ is the non-local pseudopotential operator:
\begin{eqnarray}
V_{nl} f & = & \sum_{J} V_{nl,J} f  \nonumber \\
 & = & \sum_{J} \sum_{lm} \gamma_{Jl} \left(\sum_{J'} e^{-i \bk.(\bR_J-\bR_{J'})} \chi_{J'lm} \right) \left(\sum_{J'} \int_{\Omega} \chi_{J'lm}^*(\bx,\bR_{J'}) e^{i \bk.(\bR_J-\bR_{J'})} f(\bx) \, \mathrm{d\bx} \right) \,.
\end{eqnarray}
Above, the summation index $J$ runs over all atoms in $\Omega$, and the summation index $J'$ runs over the $J^{th}$ atom and its periodic images.

The electronic ground-state is determined by solving the non-linear eigenvalue problem in Eqn. \ref{Eqn:EL} using the Self-Consistent Field (SCF) method \cite{slater1974self}. Specifically, a fixed-point iteration is performed with respect to the potential $V_{eff} = V_{xc} + \phi$, which is further accelerated using mixing/extrapolation schemes \cite{fang2009two,lin2013elliptic,pratapa2015restarted,banerjee2015periodic}. In each iteration of the SCF method, the electrostatic potential is calculated by solving the Poisson equation, and the electron density is determined by computing the eigenfunctions of the linearized Hamiltonian. The orthogonality requirement amongst the Kohn-Sham orbitals makes such a procedure scale asymptotically as $\mathcal{O}(N^3)$ with respect to the number of atoms, which severely limits the size of systems that can be studied. To overcome this restrictive scaling, $\mathcal{O}(N)$ approaches \cite{Goedecker,Bowler2012,suryanarayana2013spectral,Phanish2012Purification} will be subsequently developed and implemented into SPARC.    

\paragraph{Free energy} The free energy is evaluated using the Harris-Foulkes \cite{harris1985simplified,foulkes1989tight} type functional:
\begin{eqnarray} 
\hat{\mathcal{F}}(\bR) & = & 2\sum_{n=1}^{N_s} \fint_{BZ} g_{n}(\bk) \lambda_{n}(\bk) \, \mathrm{d\bk} + \int_{\Omega} \varepsilon_{xc} (\rho(\bx)) \rho(\bx) \, \mathrm{d \bx} - \int_{\Omega} V_{xc}(\rho(\bx))\rho(\bx)\, \mathrm{d\bx} \nonumber \\
& + & \frac{1}{2} \int_{\Omega}(b(\bx,\bR)-\rho(\bx)) \phi(\bx,\bR) \, \mathrm{d\bx} - E_{self}(\bR)  + E_c(\bR) \label{Eqn:FreeEnergy:GroundState:DFT}   \\ 
& + & 2 k_B T \sum_{n=1}^{N_{s}} \fint_{BZ} \left( g_{n}(\bk) \log g_{n}(\bk) + (1-g_{n}(\bk)) \log (1-g_{n}(\bk)) \right) \mathrm{d\bk} \,, \nonumber
\end{eqnarray}
where $E_{self}$ and $E_c$ are as defined in Eqns. \ref{Eqn:SelfEnergy} and \ref{Eqn:RepulsiveCorrectionEnergy}, respectively.

\paragraph{Atomic forces} Once the electronic ground-state has been determined, the atomic forces are calculated using the relation: 
\begin{eqnarray}
\mathbf{f}_J & = & -\frac{\partial \hat{\mathcal{F}}(\bR)}{\partial \bR_J} \nonumber \\ 
  & = &  \sum_{J'} \int_{\Omega} \nabla b_{J'}(\bx,\bR_{J'}) \left(\phi(\bx,\bR)-V_{J'}(\bx,\bR_{J'})\right) \, \mathrm{d\bx} + \mathbf{f}_{J,c}(\bR)  \label{Eqn:Force:Nuclei} \\
 & - & 4 \sum_{n=1}^{N_s} \fint_{BZ} g_{n}(\bk) \sum_{lm} \gamma_{Jl}  \Re \left[ \left( \sum_{J'} \int_{\Omega} \psi_{n}^*(\bx,\bk) e^{-i \bk .(\bR_J - \bR_{J'})} \chi_{J'lm}(\bx,\bR_{J'}) \, \mathrm{d\bx} \right) \right. \nonumber \\ 
& & \times \left. \left( \sum_{J'} \int_{\Omega} \nabla \psi_{n}(\bx,\bk) e^{i \bk .(\bR_J-\bR_{J'} )} \chi_{J'lm}^*(\bx,\bR_{J'}) \, \mathrm{d\bx} \right) \right] \, \mathrm{d \bk} \nonumber  \,,
\end{eqnarray}
where the summation index $J'$ runs over the $J^{th}$ atom and its periodic images, and  $\Re[.]$ denotes the real part of the bracketed expression. The first term is the local component of the force \cite{Phanish2012}, the second term corrects for overlapping pseudocharge densities \cite{Suryanarayana2014524} (Appendix \ref{Appendix:Correct:RepulsiveEnergy}), and the final term is the non-local component obtained by transferring the derivative on the projectors (with respect to the atomic position) to the orbitals (with respect to space) \cite{hirose2005first}. This strategy is employed since the orbitals are generally much smoother than the projectors, and therefore more accurate forces can be obtained for a given discretization \cite{pratapa2015spectral,sparc2016cluster}.\footnote{The adopted approach introduces an additional approximation into the calculation of the forces, since the derivative of the projectors can in principle be evaluated exactly, whereas the gradient operator in the finite-difference setting is an approximate one. However, we have found that this is more than offset by the smoothness of the orbitals compared to the projectors. Other approaches to improve the quality of the forces include suitable modification of the pseudopotential \cite{briggs1996real}, double-grid method \cite{OnoHir99}, and high-order spatial integration \cite{BobSchChe15}, all of which can possibly be used in conjunction with the current approach to even further reduce the egg-box effect.}  

\begin{figure}[H]
\centering
\includegraphics[keepaspectratio=true,width=0.71\textwidth]{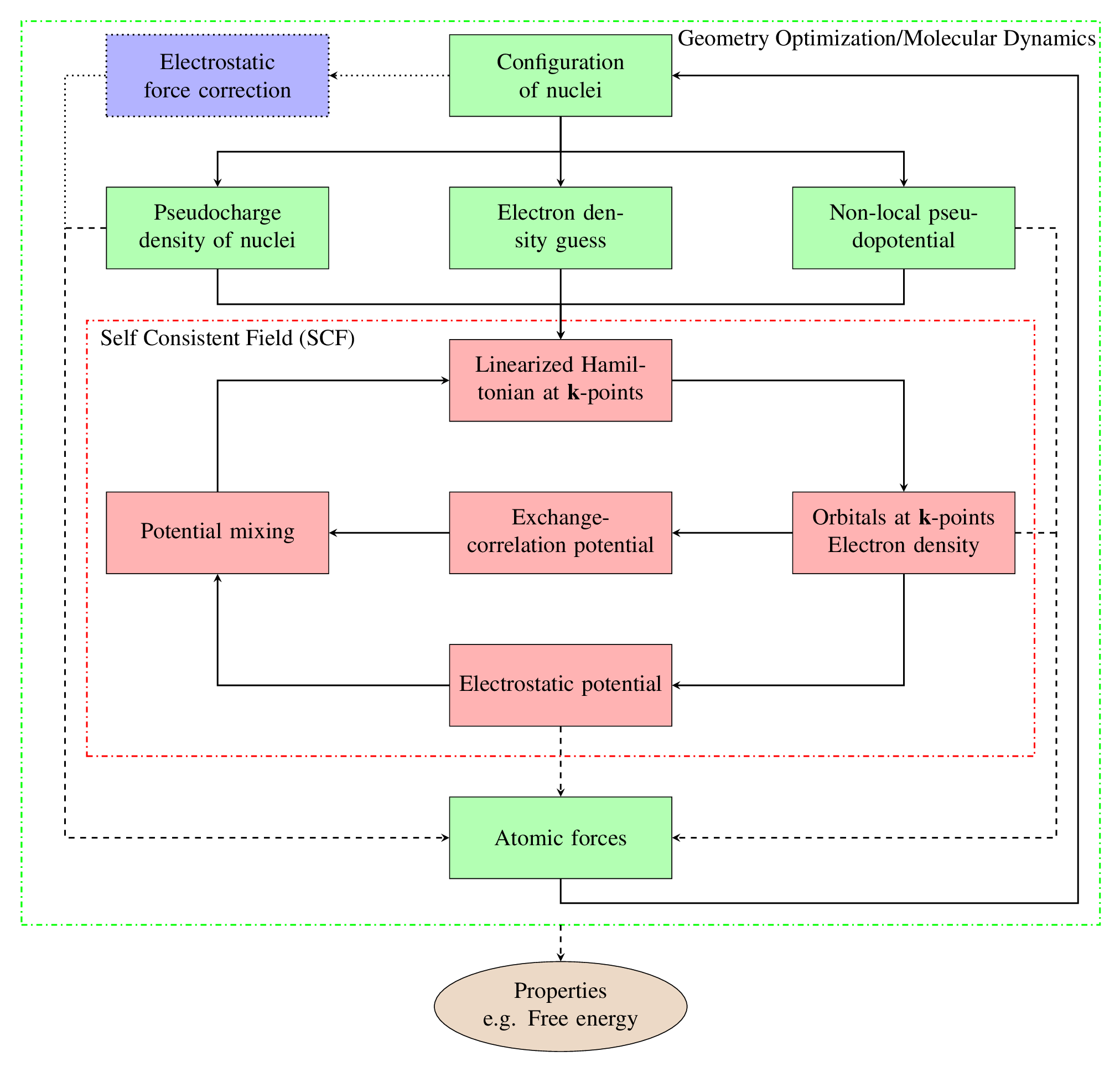} 
\caption{Outline of DFT simulations in SPARC for extended systems.}
\label{Fig:flowchart}
\end{figure}

\paragraph{Overview of SPARC} SPARC has been implemented in the framework of the Portable, Extensible Toolkit for scientific computations (PETSc) \cite{Petsc1,Petsc2} suite of data structures and routines. The approach adopted for geometry optimization/molecular dynamics is outlined in Fig. \ref{Fig:flowchart}, whose key components are discussed in detail in the subsections below. 

\subsection{Finite-difference discretization} \label{Sec:FD}
Let $\Omega$ denote the unit cell---a cuboid aligned with the $x$, $y$, and $z$ axes (origin at the center) with sides of length $L_1$, $L_2$ and $L_3$, respectively. We discretize $\Omega$ using a finite-difference grid with spacings $h_1$, $h_2$, and $h_3$ along the $x$, $y$, and $z$ axes, respectively, such that $L_1=n_1 h_1$, $L_2=n_2h_2$ and $L_3=n_3h_3$ ($n_1,n_2,n_3 \in \mathbb{N}$, $\mathbb{N}: \text{set of all natural numbers}$). We designate each finite-difference node using an index of the form $(i,j,k)$, where $i=1,2,\ldots, n_1$, $j=1,2,\ldots, n_2$ and $k=1,2,\ldots, n_3$. We approximate the Laplacian using the central finite-difference approximation:
\begin{eqnarray}\label{Eqn:FD:Laplacian}
\nabla^2_h f \big|^{(i,j,k)} \approx \sum_{p=0}^{n_o} \bigg( w_{p,1} (f^{(i+p,j,k)} + f^{(i-p,j,k)} ) + w_{p,2} ( f^{(i,j+p,k)} + f^{(i,j-p,k)} ) + w_{p,3}( f^{(i,j,k+p)} + f^{(i,j,k-p)} ) \bigg) \,,
\end{eqnarray}
where $f^{(i,j,k)}$ represents the value of the function $f$ at $(i,j,k)$, and the weights \cite{mazziotti1999spectral,ghosh2014higher}
\begin{eqnarray}
w_{0,s} & = & - \frac{1}{h_s^2} \sum_{q=1}^{n_o} \frac{1}{q^2} \,, \,\, s=1, 2, 3, \nonumber \\
w_{p,s} & = & \frac{2 (-1)^{p+1}}{h_s^2 p^2} \frac{(n_o!)^2}{(n_o-p)! (n_o+p)!} \,, \,\, p=1, 2, \ldots, n_o\,, \,\, s=1, 2, 3. \label{Eqn:weightsLap}
\end{eqnarray} 
Similarly, we approximate the gradient operator using central finite-differences:
\begin{eqnarray}\label{Eqn:gradient:approximate}
\nabla_h f \big|^{(i,j,k)} \approx  \sum_{p=1}^{n_o} \bigg( \tilde{w}_{p,1}  ( f^{(i+p,j,k)} - f^{(i-p,j,k)}) \hat{\mathbf{e}}_1 + \tilde{w}_{p,2} ( f^{(i,j+p,k)} - f^{(i,j-p,k)}) \hat{\mathbf{e}}_2  + \tilde{w}_{p,3} ( f^{(i,j,k+p)} - f^{(i,j,k-p)}) \hat{\mathbf{e}}_3 \bigg) \,,
\end{eqnarray}
where $\hat{\mathbf{e}}_1$, $\hat{\mathbf{e}}_2$ and $\hat{\mathbf{e}}_3$ represent unit vectors along the $x$, $y$, and $z$ axes, respectively. Further, the weights \cite{mazziotti1999spectral,ghosh2014higher}
\begin{equation}
\tilde{w}_{p,s} = \frac{(-1)^{p+1}}{h_s p} \frac{(n_o!)^2}{(n_o-p)! (n_o+p)!} \,, \,\, p=1, 2, \ldots, n_o\,,\,\, s=1, 2, 3.
\end{equation}
We employ the trapezoidal rule for performing spatial integrations \cite{sparc2016cluster}, i.e., 
\begin{equation} \label{Eqn:IntApprox}
\int_{\Omega} f(\bx) \, \mathrm{d\bx} \approx h_1h_2h_3  \sum_{i=1}^{n_1} \sum_{j=1}^{n_2} \sum_{k=1}^{n_3}f^{(i,j,k)} \,,
\end{equation} 
using which we approximate the nonlocal pseudopotential operator as 
\begin{eqnarray} \label{Eqn:NonlocalPS:FD}
V_{nl} f \big|^{(i,j,k)} & = & \sum_{J} V_{nl,J} f \big|^{(i,j,k)} \\
& \approx & h_1h_2h_3 \sum_{J} \sum_{lm} \sum_{p=1}^{n_1} \sum_{q=1}^{n_2} \sum_{r=1}^{n_3} \gamma_{Jl} \left( \sum_{J'} e^{-i\bk.(\bR_J-\bR_{J'})} \chi_{J'lm}^{(i,j,k)} \right) \left( \sum_{J'} e^{i\bk.(\bR_J-\bR_{J'})} \chi_{J'lm}^{*(p,q,r)} f^{(p,q,r)} \right) \nonumber \,,
\end{eqnarray}
where the summation index $J$ runs over all atoms in $\Omega$, and the summation index $J'$ runs over the $J^{th}$ atom and its periodic images. We enforce periodic boundary conditions by mapping any index that does not correspond to a node in the finite-difference grid to its periodic image within $\Omega$. We enforce zero Dirichlet boundary conditions by setting $f^{(i,j,k)}=0$ for any index that does not correspond to a node in the finite-difference grid.

Henceforth, we denote the discrete Hamiltonian matrix as a function of the Bloch wavevector $\bk$ by $\mathbf{H}(\bk) \in \C^{N_d \times N_d}$, where $N_d=n_1 \times n_2 \times n_3$ is the total number of finite-difference nodes. In addition, we represent the eigenvalues of $\mathbf{H}(\bk)$ arranged in ascending order by $\lambda_{1}(\bk), \lambda_{2}(\bk), \ldots, \lambda_{N_d}(\bk)$. We store $\mathbf{H}(\bk)$ and other sparse matrices in compressed row format, and store the discrete $u_n(\bx,\bk)$ as columns of the dense matrix $\mathbf{U}(\bk) \in \C^{N_d \times N_s}$. In parallel computations, we partition the domain as $\Omega = \bigcup\limits_{p=1}^{n_p} \Omega_p$, where $\Omega_p$ denotes the domain local to the $p^{th}$ processor, and $n_p$ is the total number of processors. The specific choice of $\Omega_p$ corresponds to the PETSc  default for structured grids.


\subsection{Pseudocharge density generation and self energy calculation} \label{Subsec:PseudochargeSelfEnergy}
In each structural relaxation/molecular dynamics step, the finite-difference Laplacian is used to assign the pseudocharge densities to the grid \cite{Phanish2012}:
\begin{equation} \label{Eqn:PseudochargeDiscrete}
b^{(i,j,k)}= \sum_{I} b_I^{(i,j,k)} \,, \quad b_I^{(i,j,k)}= - \frac{1}{4 \pi} \nabla^2_h V_I\big|^{(i,j,k)} \,,
\end{equation}
where the summation index $I$ runs over all atoms in $\R^3$. The discrete form of the pseudocharge density $b_J$ has exponential decay away from $\bR_J$ (Appendix \ref{Appendix:Pseudopotential}), which allows for its truncation at some suitably chosen radius $r_J^b$\footnote{Note that $r_{J'}^b = r_J^b$, where the index $J$ corresponds to any atom in $\Omega$, and $J'$ corresponds to the $J^{th}$ atom and its periodic images}. The corresponding discrete self energy takes the form 
\begin{equation} \label{Eqn:DiscreteEself}
E_{self}^h = \frac{1}{2} h_1h_2h_3 \sum_{I} \sum_{i=1}^{n_1} \sum_{j=1}^{n_2} \sum_{k=1}^{n_3} b_I^{(i,j,k)} V_I^{(i,j,k)} \,.
\end{equation}

In Algorithm \ref{Algo:PseudochargeCalculation}, we outline the calculation of $b^{(i,j,k)}$ and $E_{self}^h$, as implemented in SPARC. We use $P_{r^b}^p$ to denote the set of all atoms for which $\Omega_{r_{J'}^b} \cap \Omega_p \neq \emptyset$, where the index $J'$ runs over all atoms in $\Omega$ and their periodic images. Here, $\Omega_{r^b_{J'}}$ represents the cuboid with side of lengths $2r_{J,1}^b \approx 2r_{J,2}^b \approx 2r_{J,3}^b$ ($\approx 2 r_{J}^b$) centered on the ${J'}^{th}$ atom. We have chosen $\Omega_{r^b_{J'}}$ to be a cuboid rather than a sphere due to its simplicity and efficiency within the Euclidean finite-difference discretization. The values of $r_{J,1}^b$, $r_{J,2}^b$, and $r_{J,3}^b$---integer multiples of the mesh spacings $h_1$, $h_2$, and $h_3$, respectively---are chosen such that the charge constraint in Eqn. \ref{Eqn:IntegralPseudocharges} is satisfied to within a prespecified tolerance $\varepsilon_b$, i.e.,
\begin{equation} \label{Eqn:PseudochargeRadiusChoice}
\left| \frac{h_1h_2h_3 \sum_{J'} \sum_{i=1}^{n_1} \sum_{j=1}^{n_2} \sum_{k=1}^{n_3} b_{J'}^{(i,j,k)} - Z_J }{Z_J} \right| < \varepsilon_b \,. 
\end{equation} 
For the atom indexed by $J'$ belonging to $P_{r^b}^p$, we determine the overlap region $\Omega_{r^b_{J'}} \cap \Omega_p$, with the subscripts $s$ and $e$ denoting the starting and ending indices, respectively. In this overlap region (and an additional $2 n_0$ points in each direction), we interpolate the values of $V_{J'}^{(i,j,k)}$ on to the finite-difference grid using cubic-splines \cite{ahlberg1967theory}. Next, we determine $b^{(i,j,k)}$ and $E_{self}^{h,p}$---contribution of the $p^{th}$ processor to the self energy---using Eqns. \ref{Eqn:PseudochargeDiscrete} and \ref{Eqn:DiscreteEself}, respectively. Finally, we calculate the total self energy $E_{self}^h$ by summing the contributions from all the processors. 

\begin{algorithm}[H] \label{Algo:PseudochargeCalculation}
{\bf{Input}}: $\bR$, $V_{J}$, and $r_J^b$ \\
$b^{(i,j,k)}=0$, $E_{self}^{h,p}=0$\\
\For{$J' \in P_{r^b}^p$}
{
Determine $i_{s}$, $i_e$, $j_{s}$, $j_e$, $k_{s}$, $k_e$ of $\Omega_{r^b_{J'}} \cap \Omega_p$ \\
Determine $V_{J'}^{(i,j,k)}$ $\forall$ $i \in [i_s-n_o, i_e+n_o]$, $j \in [j_s - n_o, j_e+n_o]$, $k \in [k_s - n_o, k_e+n_o]$ \\
$b^{(i,j,k)}_{J'} = - \frac{1}{4\pi} \nabla^2_h V_{J'} \big|^{(i,j,k)}$, \quad $b^{(i,j,k)} = b^{(i,j,k)} + b^{(i,j,k)}_{J'} $ $\forall$ $i \in [i_s,i_e]$, $j \in [j_s,j_e]$, $k \in [k_s,k_e]$ \\
$E_{self}^{h,p} = E_{self}^{h,p} + \frac{1}{2} h_1h_2h_3 b^{(i,j,k)}_{J'} V_{J'}^{(i,j,k)}$ $\forall$ $i \in [i_s,i_e]$, $j \in [j_s,j_e]$, $k \in [k_s,k_e]$ \\  
}
$E_{self}^h = \sum_{p=1}^{n_p} E_{self}^{h,p}$ \\ 
{\bf{Output}}: $b^{(i,j,k)}$ and $E_{self}^h$
\caption{Pseudocharge density generation and self energy calculation}
\end{algorithm} 

\subsection{Brillouin zone integration} \label{Subsec:BZ:Int}
The volume averaged integral of any function over the Brillouin zone is approximated as
\begin{equation} \label{Eqn:BZ:Integration}
\fint_{BZ} f(\bk) \, \mathrm{d\bk} \approx \sum_{b=1}^{N_k} w_b f_b \,,
\end{equation}
where $f_b \equiv f(\bkb)$. Here, $\bkb$ and $w_b$ ($b=1,2, \ldots N_k$) denote the nodes and weights for integration, respectively. The specific choice of $\bkb$ is commonly referred to as Brillouin zone sampling. 

\subsection{Electron density calculation} \label{Subsec:Rho}
On employing the Brillouin zone integration scheme described by Eqn. \ref{Eqn:BZ:Integration}, the electron density takes the form
\begin{equation}
\rho(\bx) = 2 \sum_{n=1}^{N_s} \sum_{b=1}^{N_k} w_b g_{nb} | u_{nb}(\bx) |^2 \,.
\end{equation}
In each SCF iteration, we calculate the electron density using the Chebyshev-filtered subspace iteration (CheFSI) method \cite{zhou2006self,zhou2006parallel}, wherein we compute approximations to the eigenvectors corresponding to the lowest $N_s$ eigenvalues of $\mathbf{H_b}, \, b=1, 2, \ldots N_k$. This choice of eigensolver is motivated by the minimal orthogonalization and computer memory costs compared to other alternatives commonly employed in electronic structure calculations, e.g. Locally Optimal Block Preconditioned Conjugate Gradient (LOBPCG) \cite{knyazev2001toward}. Additionally, it has been shown to be extremely efficient within SPARC for isolated systems \cite{sparc2016cluster}.

The implementation of the CheFSI algorithm in SPARC for extended systems consists of three main components. First, we use the rapid growth of Chebyshev polynomials outside the interval $[-1,1]$ to filter out the unwanted eigencomponents from $\mathbf{U_b}$: 
\begin{equation} \label{Eqn:ChebFilter}
\mathbf{U_{bf}} = p_{mb}(\mathbf{H_{b}}) \mathbf{U_{b}} \,, \quad p_{mb}(t) = C_m \left( \frac{t-c_b}{e_b} \right) \,, \quad b=1, 2, \ldots, N_k \,,
\end{equation}
where the columns of $\mathbf{U_{bf}}$ represent the filtered $\mathbf{U_{b}}$, and $C_m$ denotes the Chebyshev polynomial of degree $m$. Additionally, $e_b = (\lambda_{N_d b} - \lambda_c)/2$ and $c_b = (\lambda_{N_d b} + \lambda_c)/2$, where $\lambda_c$ is the filter cutoff. Rather than explicitly compute the matrix $p_{mb}(\mathbf{H_{b}})$, its product with $\mathbf{U_{b}}$ is determined using the three term recurrence relation for Chebyshev polynomials.

Next, projecting onto the filtered basis ${\mathbf{U_{bf}}}$, we arrive at the generalized eigenproblem: 
\begin{equation}\label{Eqn:SubspaceGeneralizedEigenproblem}
\mathbf{\tilde{H}_{b}} \mathbf{y_{nb}} = \lambda_{nb} \mathbf{\tilde{M}_{b}} \mathbf{y_{nb}} \,, \quad  n=1,2, \ldots N_s \,,\quad b=1, 2, \ldots, N_k \,,
\end{equation}
where $\lambda_{nb}$ represent approximations to the eigenvalues of $\mathbf{H_{b}}$, and the dense matrices $\mathbf{\tilde{H}_{b}}, \mathbf{\tilde{M}_{b}} \in \C^{N_s \times N_s}$ are determined as
\begin{equation}
\mathbf{\tilde{H}_{b}} = \mathbf{U_{bf}^{*T}} \mathbf{H_{b}} \mathbf{U_{bf}} \,, \quad \mathbf{\tilde{M}_{b}} = \mathbf{U_{bf}^{*T}} \mathbf{U_{bf}} \,.
\end{equation}
After eigendecomposing Eqn. \ref{Eqn:SubspaceGeneralizedEigenproblem} at all the \textbf{k}-points, the Fermi energy $\lambda_f$ is calculated by enforcing the constraint on the total number of electrons:
\begin{equation}
2 \sum_{n=1}^{N_s} \sum_{b=1}^{N_k} w_{b} g_{nb} = N_e \,, \quad \text{where} \quad g_{nb} = \left(1 + \exp\left( \frac{\lambda_{nb} - \lambda_f}{k_B T} \right) \right)^{-1} \,.
\end{equation}

Finally, we perform the subspace rotation 
\begin{equation}\label{Eqn:OrbitalTransformation}
\mathbf{U_b} = \mathbf{U_{bf}} \mathbf{Y_b} \,,  \quad b=1, 2, \ldots, N_k \,,
\end{equation}  
where the columns of the matrix $\mathbf{Y_b} \in \C^{N_s \times N_s}$ contain the eigenvectors $\mathbf{y_{nb}}$. The columns of $\mathbf{U_b}$ so obtained represent approximations to the eigenvectors of $\mathbf{H_b}$. The electron density at the finite-difference grid points is then calculated using the relation
\begin{equation}
\rho^{(i,j,k)} =  2 \displaystyle\sum_{n=1}^{N_s} \sum_{b=1}^{N_k} w_b g_{nb} |u_{nb}^{(i,j,k)}|^2 \,,
\end{equation}
where the values of $u_{nb}^{(i,j,k)}$ are extracted from the $n^{th}$ column of $\mathbf{U_b}$.

We start the very first SCF iteration of the complete DFT simulation with a randomly generated guess for $\mathbf{U_b}$ ($b=1, 2, \ldots, N_k$), and perform the CheFSI steps multiple times \cite{zhou2014chebyshev} without calculating/updating the electron density. This allows us to obtain a good approximation of the electron density for the second SCF iteration. For every subsequent SCF iteration, we perform the CheFSI steps only once, and use the rotated $\mathbf{U_b}$ from the previous step as the initial guess. Overall, the calculation of the electron density scales as $\mathcal{O}(N_k N_s N_d) + \mathcal{O}(N_k N_s^2 N_d) + \mathcal{O}(N_k N_s^3)$, which makes it $\mathcal{O}(N^3)$ with respect to the number of atoms.

\subsection{Free energy calculation}
On approximating the spatial integrals using Eqn. \ref{Eqn:IntApprox} and the Brillouin zone integrals using Eqn. \ref{Eqn:BZ:Integration}, we arrive at the discrete form for the free energy at the electronic ground-state: 
\begin{eqnarray} 
\hat{\mathcal{F}}^{h} & = & 2\sum_{n=1}^{N_s} \sum_{b=1}^{N_k} w_b g_{nb} \lambda_{nb} +  h_1h_2h_3  \sum_{i=1}^{n_1} \sum_{j=1}^{n_2} \sum_{k=1}^{n_3} \bigg( \varepsilon_{xc}^{(i,j,k)}  \rho^{(i,j,k)} - V_{xc}^{(i,j,k)}\rho^{(i,j,k)} + \frac{1}{2} (b^{(i,j,k)}-\rho^{(i,j,k)}) \phi^{(i,j,k)} \bigg) \nonumber \\
& & - E_{self}^h + E_c^h + 2 k_B T \sum_{n=1}^{N_{s}} \sum_{b=1}^{N_k} w_b \left( g_{nb} \log g_{nb} + (1-g_{nb}) \log (1-g_{nb}) \right)   \label{Eqn:FreeEnergy:GroundState:DFT:discrete} \,,
\end{eqnarray} 
where $E_{self}^h$ is the discrete self energy (Eqn.~\ref{Eqn:DiscreteEself}), and $E_c^h$ is the discrete repulsive energy correction for overlapping pseudocharges (Eqn.~\ref{Eqn:Ec:Disc}). The evaluation of $\hat{\mathcal{F}}^{h}$ scales as $\mathcal{O}(N_d)$, and therefore $\mathcal{O}(N)$ with respect to the number of atoms.


\subsection{Atomic forces calculation}
The discrete form of the atomic force presented in Eqn. \ref{Eqn:Force:Nuclei} can be split into three parts:
\begin{equation}
\mathbf{f}_J^h = \mathbf{f}_{J,loc}^h + \mathbf{f}_{J,c}^h + \mathbf{f}_{J,nloc}^h \,,
\end{equation}
where $\mathbf{f}_{J,loc}^h$ is the discrete local component, $\mathbf{f}_{J,c}^h$ is the discrete correction corresponding to overlapping pseudocharges, and $\mathbf{f}_{J,nloc}^h$ is the discrete non-local component of the force. Below, we discuss the evaluation of $\mathbf{f}_{J,loc}^h$ and $\mathbf{f}_{J,nloc}^h$ in SPARC for extended systems, with the computation of $\mathbf{f}_{J,c}^h$ proceeding similarly to $\mathbf{f}_{J,loc}^h$. 

\paragraph{Local component} On approximating the spatial integral using Eqn. \ref{Eqn:IntApprox}, the discrete local component of the force takes the form
\begin{equation} \label{Eqn:LocalForce:Discrete}
\mathbf{f}_{J,loc}^h = h_1h_2h_3  \sum_{J'} \sum_{i=1}^{n_1} \sum_{j=1}^{n_2} \sum_{k=1}^{n_3} \mathbf{\nabla}_h b_{J'}\big|^{(i,j,k)} (\phi^{(i,j,k)} - V_{J'}^{(i,j,k)}) \,,
\end{equation}
where the summation index $J'$ runs over the $J^{th}$ atom and its periodic images. The calculation of $\mathbf{f}_{J,loc}^h$ is outlined in Algorithm \ref{Algo:LocalForceCalculation}, which proceeds as follows. For the atom indexed by $J'$ belonging to $P_{r^b}^p$, we determine the overlap region $\Omega_{r^b_{J'}} \cap \Omega_p$. In this overlap region (and an additional $4n_0$ points in each direction), the values of $V_{J'}^{(i,j,k)}$ are interpolated on to the finite-difference grid using cubic-splines, from which $b_{J'}^{(i,j,k)}$ is calculated using Eqn. \ref{Eqn:PseudochargeDiscrete}. Subsequently, $\mathbf{f}_{J,loc}^{h,p}$---contribution of the $p^{th}$ processor to the local component of the force---is calculated using Eqn. \ref{Eqn:LocalForce:Discrete}. Finally, the contributions from all processors are summed to obtain $\mathbf{f}_{J,loc}^h$. 

\begin{algorithm}[H] \label{Algo:LocalForceCalculation}
\textbf{Input}: $\bR$, $\phi^{(i,j,k)}$, $V_J$, and $r_J^b$ \\
$\mathbf{f}_{J,loc}^{h,p} =0$ \\
\For{$J' \in P_{r^b}^p$}
{
Determine $i_s$, $i_e$, $j_s$, $j_e$, $k_s$, $k_e$ of $\Omega_{r_{J'}^b} \cap \Omega_p$ \\
Determine $V_{J'}^{(i,j,k)}$ $\forall$ $i \in [i_s-2n_o, i_e+2n_o]$, $j \in [j_s - 2n_o, j_e+2n_o]$, and $k \in [k_s - 2n_o, k_e+2n_o]$ \\
$b_{J'}^{(i,j,k)} = - \frac{1}{4\pi} \nabla^2_h V_{J'}\big|^{(i,j,k)}$ $\forall$ $i \in [i_s-n_0,i_e+n_0]$, $j \in [j_s-n_0,j_e+n_0]$, $k \in [k_s-n_0,k_e+n_0]$ \\  
$\mathbf{f}_{J,loc}^{h,p} = \mathbf{f}_{J,loc}^{h,p} + h_1 h_2 h_3 \sum_{i=i_s}^{i_e} \sum_{j=j_s}^{j_e} \sum_{k=k_s}^{k_e} \nabla_h b_{J'} \big|^{(i,j,k)} ( \phi^{(i,j,k)} - V_{J'}^{(i,j,k)} ) $  \\
} 

$\mathbf{f}_{J,loc}^h = \sum_{p=1}^{n_p} \mathbf{f}_{J,loc}^{h,p}$ \\
\textbf{Output}: $\mathbf{f}_{J,loc}^h$
\caption{Calculation of the local component of the atomic force.}
\end{algorithm}

\paragraph{Non-local component} On approximating the spatial integral using Eqn. \ref{Eqn:IntApprox}, the discrete non-local component of the force takes the form 
\begin{equation} \label{Eqn:NonlocForce:Discrete}
\mathbf{f}_{J,nloc}^h  =  -4\sum_{n=1}^{N_s} \sum_{b=1}^{N_k} w_b g_{nb} \sum_{lm} \gamma_{Jl} \Re \left[ Y_{Jnblm} \mathbf{W}_{Jnblm} \right]  \,,
\end{equation}
where 
\begin{eqnarray}
Y_{Jnblm} & = & h_1 h_2 h_3  \sum_{J'} \sum_{i=1}^{n_1} \sum_{j=1}^{n_2} \sum_{k=1}^{n_3} \psi_n^{*(i,j,k)} e^{-i \bkb.(\bR_J-\bR_{J'})} \chi_{J'lm}^{(i,j,k)} \,, \label{Eqn:NonLocForce:U} \\
\mathbf{W}_{Jnblm} & = & h_1 h_2 h_3 \sum_{J'} \sum_{i=1}^{n_1} \sum_{j=1}^{n_2} \sum_{k=1}^{n_3} \nabla_h \psi_n \big|^{(i,j,k)} e^{i \bkb.(\bR_J-\bR_{J'})} \chi_{J'lm}^{*(i,j,k)} \,. \label{Eqn:NonLocForce:W}
\end{eqnarray}

In Algorithm \ref{Algo:non-localForceCalculation}, we outline the calculation of $\mathbf{f}_{J,nloc}^h$. We use $P_{r^c}^p$ to denote the set of all atoms for which $\Omega_{r_{J'}^c} \cap \Omega_p \neq \emptyset$, where the index $J'$ runs over all atoms in $\Omega$ and their periodic images. In addition, $\Omega_{r^c_{J'}}$ represents the cuboid with side of lengths $2r_{J,1}^c \approx 2r_{J,2}^c \approx 2r_{J,3}^c (\approx 2r_{J}^c$)---integer multiples of the spacings $h_1$, $h_2$, and $h_3$, respectively---centered on the ${J'}^{th}$ atom. We have chosen $\Omega_{r^c_{J'}}$ to be a cuboid rather than a sphere due to its simplicity and efficiency within the Euclidean finite-difference discretization. For the atom indexed by $J'$ belonging to $P_{r^c}^p$, we determine the overlap region $\Omega_{r^c_{J'}} \cap \Omega_p$, with the subscripts $s$ and $e$ denoting the starting and ending indices, respectively. In this overlap region, we interpolate the radial components of the projectors $\chi_{J'lm}^{(i,j,k)}$ on to the finite-difference grid using cubic-splines. Next, we utilize Eqns. \ref{Eqn:NonLocForce:U} and \ref{Eqn:NonLocForce:W} to determine $Y_{Jnlm}^p$ and $\mathbf{W}_{Jnlm}^p$, respectively, which represent the contributions of the $p^{th}$ processor to $Y_{Jnlm}$ and $\mathbf{W}_{Jnlm}$, respectively. Finally, we sum the contributions from all the processors to obtain $Y_{Jnlm}$ and $\mathbf{W}_{Jnlm}$, which are then used to calculate $\mathbf{f}_{J,nloc}^h$ using Eqn. \ref{Eqn:NonlocForce:Discrete}. Overall, the calculation of the atomic forces scales as $\mathcal{O}(N)$ with respect to the number of atoms.

\begin{algorithm}[H] \label{Algo:non-localForceCalculation}
\textbf{Input}: $\bR$, $\psi_n^{(i,j,k)}$, $\gamma_{Jl}$, $\chi_{Jlm}$, and $r_J^c$ \\
$Y_{Jnblm}^p=0$, $\mathbf{W}_{Jnblm}^p=0$ \\

\For{$J' \in P_{r^c}^p$} 
{
 Determine starting and ending indices $i_s$, $i_e$, $j_s$, $j_e$, $k_s$, $k_e$ for $\Omega_{r_{J'}^c} \cap \Omega_p$ \\
 Determine $\chi_{J'lm}^{(i,j,k)}$ $\forall$ $i \in [i_s,i_e]$, $j \in [j_s,j_e]$, $k \in [k_s,k_e]$ \\  
 $Y_{Jnblm}^p = Y_{Jnblm}^p + h_1 h_2 h_3 \psi_n^{*(i,j,k)} e^{-i \bkb.(\bR_J-\bR_{J'})} \chi_{J'lm}^{(i,j,k)}$ $\forall$ $i \in [i_s,i_e]$, $j \in [j_s,j_e]$, $k \in [k_s,k_e]$ \\
$\mathbf{W}_{Jnblm}^p = \mathbf{W}_{Jnblm}^p + h_1 h_2 h_3 \nabla_h \psi_n \big|^{(i,j,k)} e^{i \bkb.(\bR_J-\bR_{J'})} \chi_{J'lm}^{*(i,j,k)}$ $\forall$ $i \in [i_s,i_e]$, $j \in [j_s,j_e]$, $k \in [k_s,k_e]$ \\
}

$Y_{Jnblm} = \sum_{p=1}^{n_p} Y_{Jnblm}^p$, $\mathbf{W}_{Jnblm} = \sum_{p=1}^{n_p} \mathbf{W}_{Jnblm}^p$ \\ 
$\mathbf{f}_{J,nloc}^h  =  -4\sum_{n=1}^{N_s} \sum_{b=1}^{N_k} w_b g_{nb} \sum_{lm} \gamma_{Jl} \Re \left[ Y_{Jnblm} \mathbf{W}_{Jnblm} \right] $ \\
\textbf{Output}: $\mathbf{f}_{J,nloc}^h$
\caption{Calculation of the non-local component of the atomic force }
\end{algorithm} 


\section{Examples and Results} \label{Sec:Examples}
In this section, we verify the accuracy and efficiency of SPARC for performing DFT calculations of extended systems. In all the examples, we utilize a twelfth-order accurate finite-difference discretization ($n_0=6$), the Perdew-Wang parametrization \cite{perdew1992accurate} of the correlation energy calculated by Ceperley-Alder \cite{Ceperley1980}, norm-conserving Troullier-Martins pseudopotentials \cite{Troullier}, and a smearing of $k_B T = 0.01$ Ha. The cutoff radii for the non-local projectors and the local component of the pseudopotentials are listed in Appendix \ref{Appendix:Pseudopotential}. We utilize the Monkhorst-Pack \cite{monkhorst1976special} grid for performing integrations over the Brillouin zone.  Unless specified otherwise, the simulations correspond to $\bk=$[$0.0$ $0.0$ $0.0$] ($\Gamma$-point). We use the notation $h$ to denote the mesh-size when a common spacing is employed in all directions, i.e., $h_1=h_2=h_3=h$.

We solve the linear system corresponding to the Poisson problem using the block-Jacobi preconditioned \cite{golub2012matrix} minimal residual method (MINRES) \cite{paige1975solution}. Within the CheFSI approach, we utilize a polynomial of degree $m=20$ for Chebyshev filtering; the Lanczos method \cite{lanczos1950iteration} for calculating the extremal eigenvalues of the Hamiltonian $\mathbf{H_b}$; and LAPACK's \cite{laug} implementation of the QR algorithm \cite{watkins2004fundamentals} for solving the subspace eigenproblem in Eqn. \ref{Eqn:SubspaceGeneralizedEigenproblem}. We calculate the Fermi energy using Brent's method \cite{press2007numerical}, and use Anderson extrapolation \cite{anderson1965iterative} with relaxation parameter of $0.3$ and mixing history of $7$ for accelerating the convergence of the SCF method. We employ the Polak-Ribiere variant of non-linear conjugate gradients with a secant line search \cite{Shewchuk1994} for geometry optimization. Finally, we use the leapfrog integration scheme \cite{Rapaport} for performing NVE molecular dynamics. 

All the results presented here are converged to within the `chemical accuracy' of $0.001$ Ha/atom in the energy and $0.001$ Ha/Bohr in the forces.\footnote{In this section, we will use the term `chemical accuracy' to denote the convergence of energy and atomic forces---with respect to the fully converged DFT results---to within $0.001$ Ha/atom and $0.001$ Ha/Bohr, respectively.} Wherever applicable, the results obtained by SPARC are compared to the well-established plane-wave code ABINIT \cite{ABINIT,gonze2009abinit,gonze2005brief}. The error in energy is defined to be the difference in the magnitude, and the error in forces is defined to be the maximum difference in any component. All simulations are performed on a  computer cluster consisting of $16$ nodes with the following configuration: Altus 1804i Server - 4P Interlagos Node, Quad AMD Opteron 6276, 16C, 2.3 GHz, 128GB, DDR3-1333 ECC, 80GB SSD, MLC, 2.5" HCA, Mellanox ConnectX 2, 1-port QSFP, QDR, memfree, CentOS, Version 5, and connected through InfiniBand cable.


\subsection{Convergence with discretization} 
First, we verify convergence of the energy and atomic forces computed by SPARC with respect to spatial discretization. For this study, we choose three examples: $2\times 2\times 2$ unit cells of lithium hydride with lattice constant of $7.37$ Bohr and corner lithium atom perturbed by [$0.57$ $0.43$ $0.37$] Bohr; $2\times 2\times 2$ unit cells of silicon with lattice constant of $10.68$ Bohr and corner atom perturbed by [$0.93$ $0.50$ $0.20$] Bohr; and $2\times 2\times 2$ unit cells of gold with lattice constant of $8.0$ Bohr, $1\times 1 \times 2$ Brillouin zone integration, and a face-centered atom perturbed by [$0.85$ $0.50$ $0.30$] Bohr. All errors are defined with respect to ABINIT, wherein we employ plane-wave cutoffs of $46$, $40$, and $46$ Ha for lithium hydride, silicon, and gold, respectively. This results in energy and forces that are converged to within $5\times 10^{-6}$ Ha/atom and $5\times 10^{-6}$ Ha/Bohr, respectively. 

\begin{figure}[H]
\centering
\subfloat[Energy]{\label{fig:energyConvergence}\includegraphics[keepaspectratio=true,width=0.48\textwidth]{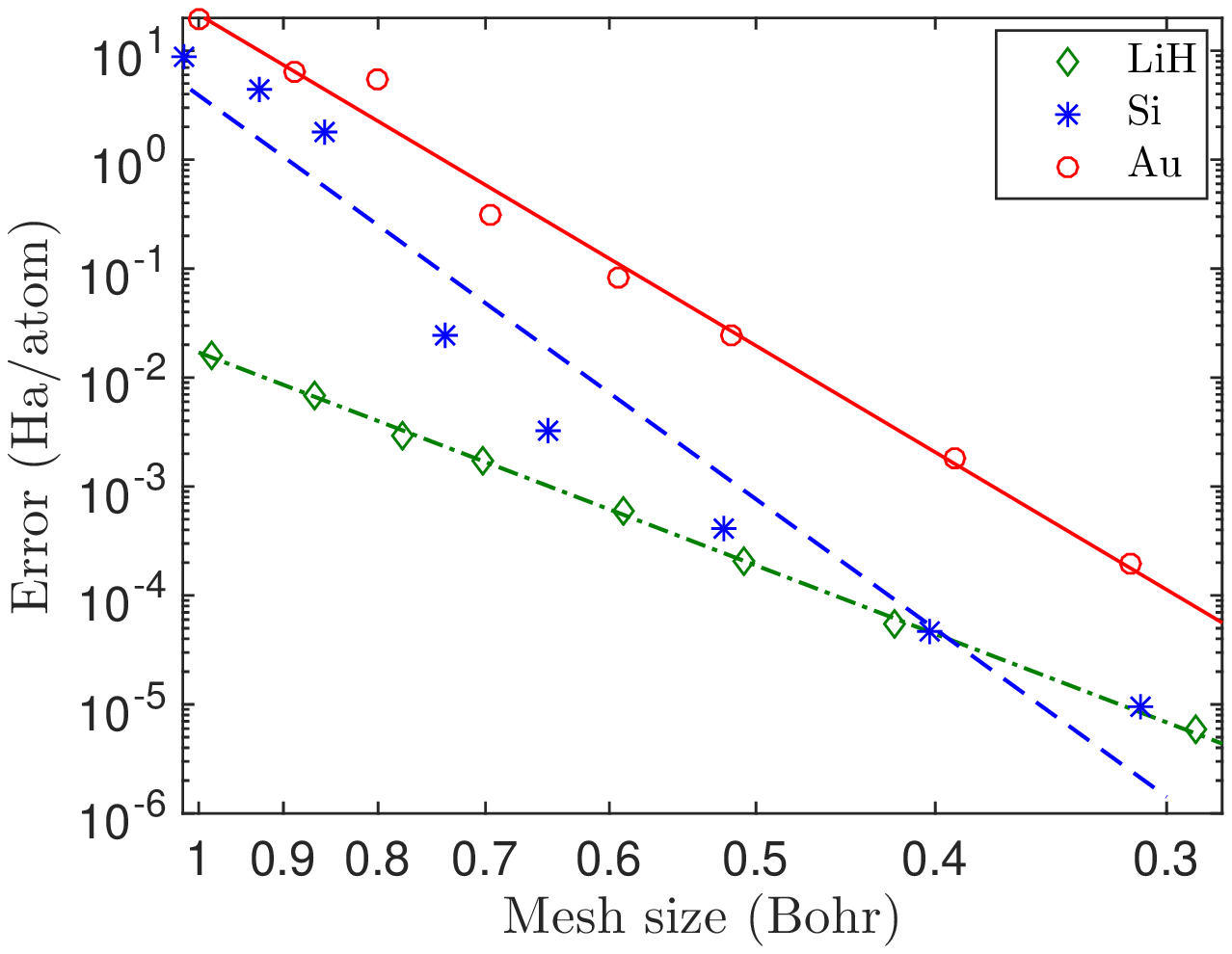}}
\subfloat[Forces]{\label{fig:forceConvergence}\includegraphics[keepaspectratio=true,width=0.48\textwidth]{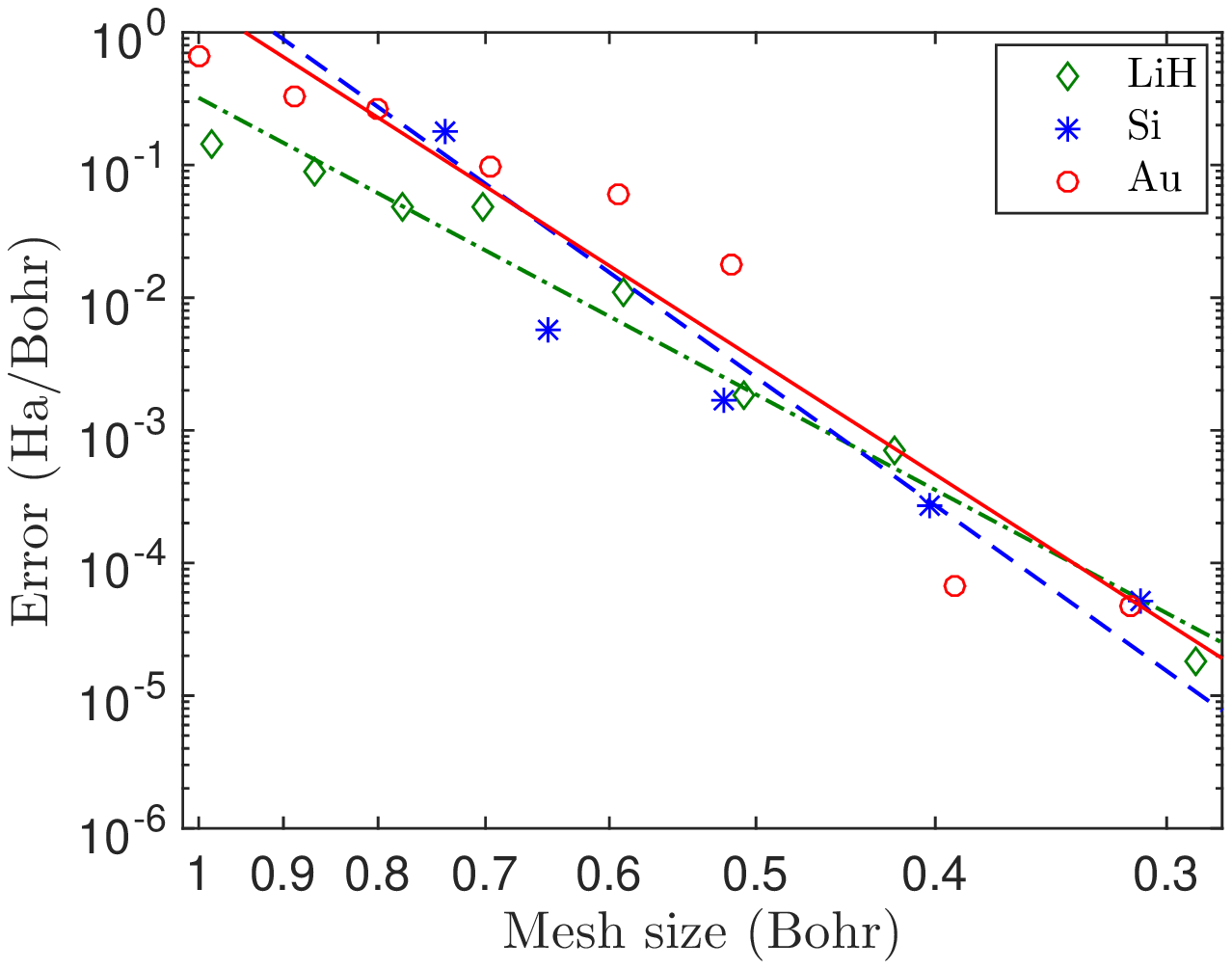}}
\caption{Convergence of the energy and atomic forces with respect to mesh size to reference planewave result for the lithium hydride, silicon, and gold systems. The straight lines represent linear fits to the data.}
\label{fig:convergenceDiscretization}
\end{figure}

It is clear from Fig. \ref{fig:convergenceDiscretization}---plots of the error in energy and atomic forces with respect to mesh-size---that there is systematic convergence to the reference plane-wave result. On performing a fit to the data, we obtain average convergence rates of approximately $\mathcal{O}(h^{10})$ in the energy and $~\mathcal{O}(h^{9})$ in the forces. In doing so, the chemical accuracy desired in electronic structure calculations is readily attained. These results demonstrate that SPARC is able to obtain high convergence rates in both the energy and forces, which contributes to its accuracy and efficiency. Moreover, the energies and forces in SPARC converge at comparable rates, without the need for additional measures such as double-grid \cite{OnoHir99} or high-order integration \cite{BobSchChe15} techniques.\footnote{Since the calculation of the forces involves taking derivatives of the energy with respect to the atomic positions, they are susceptible to larger egg-box effects, which contributes to a typically slower convergence of the forces relative to the energy \cite{BobSchChe15}.} 

\subsection{Convergence with vacuum size for slabs and wires}\label{Subsection:Convergence:Vacuum}
Next, we verify the convergence of the energy and atomic forces computed by SPARC with respect to vacuum size for slabs and wires. As representative examples, we choose a $1 \times 1 \times 5$ aluminum slab with lattice constant $7.78$ Bohr; and a silicon nanowire of lattice constant $10.16$ Bohr, radius $15$ Bohr and length $10.16$ Bohr. The slab has vacuum in the $z$ direction, and the wire has vacuum in the $y$ and $z$ directions. We utilize mesh-spacings of $\{h_1, h_2, h_3 \} = \{ 0.5985, 0.5985, 0.6 \}$ Bohr and $\{h_1, h_2, h_3 \} = \{ 0.406, 0.4, 0.4 \}$ Bohr for the slab and wire, respectively.  

\begin{figure}[H]
\centering
\subfloat[Aluminum slab]{\label{Fig:PartialPeriodic:ConvergenceDomain:AlSlab} \includegraphics[keepaspectratio=true,width=0.48\textwidth]{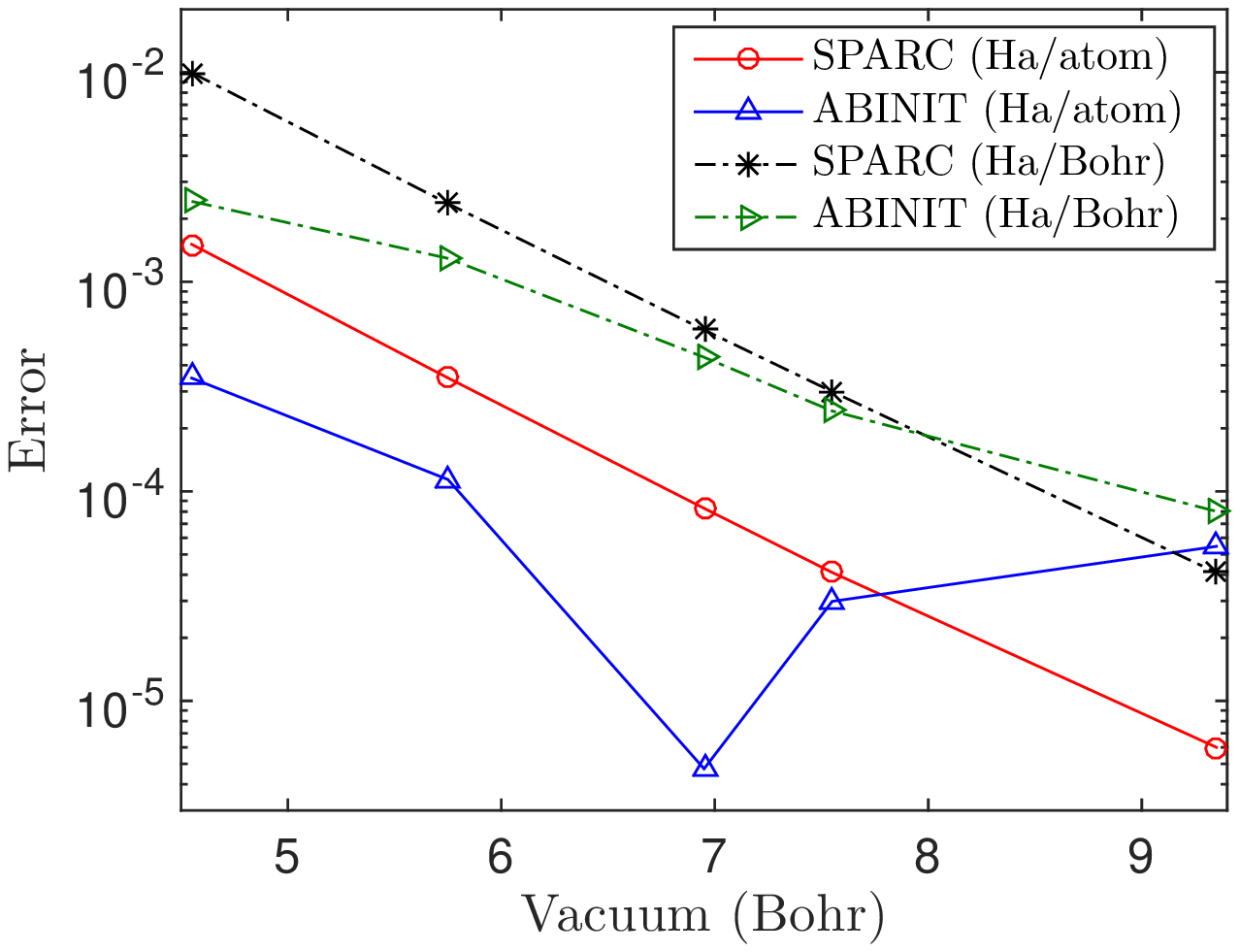}} 
\subfloat[Silicon nanowire]{\label{Fig:PartialPeriodic:ConvergenceDomain:SiWire} \includegraphics[keepaspectratio=true,width=0.48\textwidth]{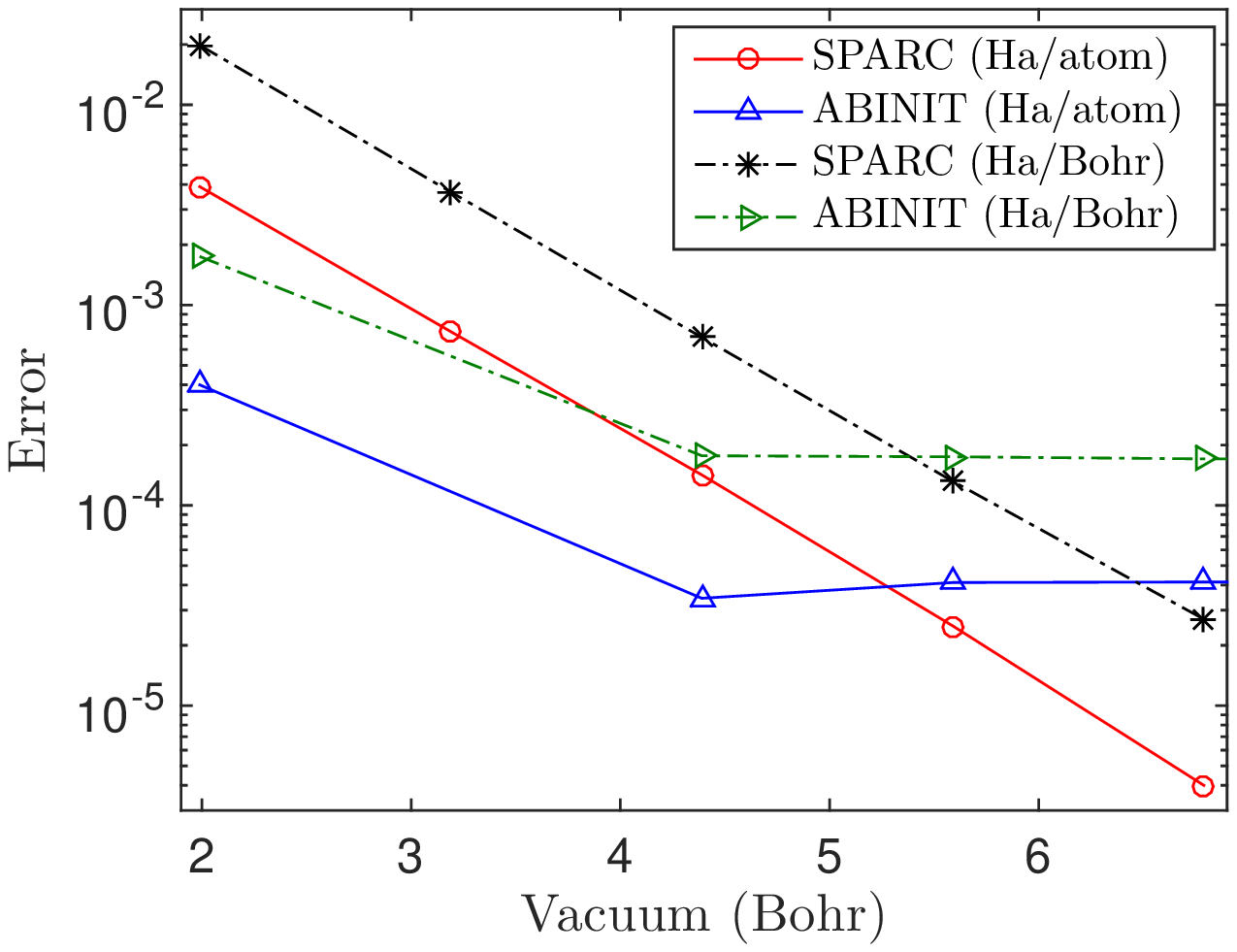}} 
\caption{Convergence of energy and atomic forces with respect to vacuum size for the aluminum slab and silicon wire.}
\label{Fig:ConvergenceDomain:comparison}
\end{figure}

In Fig. \ref{Fig:ConvergenceDomain:comparison}, we present convergence of the energy and atomic forces with vacuum size for SPARC and ABINIT, wherein the results obtained for a vacuum of $18$ Bohr are used as reference. We observe that SPARC achieves exponential convergence in both the energy and forces to accuracies well below those desired in DFT calculations. In particular, even a vacuum of around $7$ Bohr is sufficient to obtain chemical accuracy in both energy and forces.\footnote{Indeed, the convergence with vacuum size can be further accelerated in SPARC (particularly for systems with large dipole moments) by incorporating more accurate boundary conditions for slabs and wires \cite{natan2008real}.} However, in ABINIT, there is a stagnation of the error, which can be attributed to the spurious interactions between images resulting from the need for periodic boundary conditions. 

\subsection{Bulk properties}
\begin{figure}[H]
\centering
\includegraphics[keepaspectratio=true,width=0.48\textwidth]{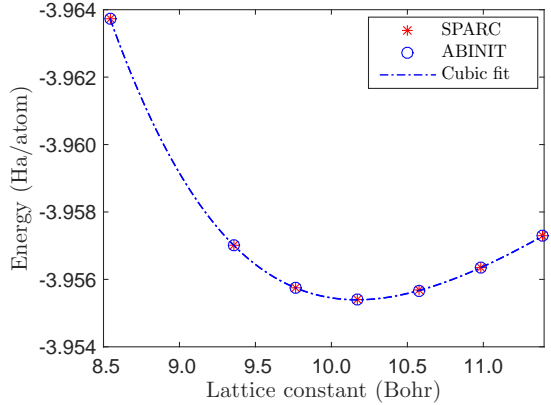}
\caption{Variation of energy with lattice constant for bulk silicon.}
\label{fig:EnergyLattice}
\end{figure}

We now verify the ability of SPARC to accurately calculate material bulk properties. We select silicon---$8$-atom unit cell with $4\times 4\times 4$ Brillouin zone integration---as the representative example. In SPARC, we use a mesh-size of $h = 0.407$ Bohr. We compare the results with ABINIT, wherein we choose a plane-wave energy cutoff of $40$ Ha, which results in energies that are converged to within $5 \times 10^{-6}$ Ha/atom. In Fig. \ref{fig:EnergyLattice}, we plot the energy so computed by SPARC and ABINIT as a function of the lattice constant. We observe that there is very good agreement between SPARC and ABINIT, with the curves being practically indistinguishable. From a cubic fit to the data, we find that the predicted equilibrium lattice constant, energy, and bulk modulus are in agreement to within $0.003$ Bohr, $1 \times 10^{-5}$ Ha/atom, and $0.18$ GPa, respectively.  

\begin{figure}[H]
\centering
\includegraphics[trim =0cm 1.2cm 0cm 0cm,clip,keepaspectratio=true,width=0.48\textwidth]{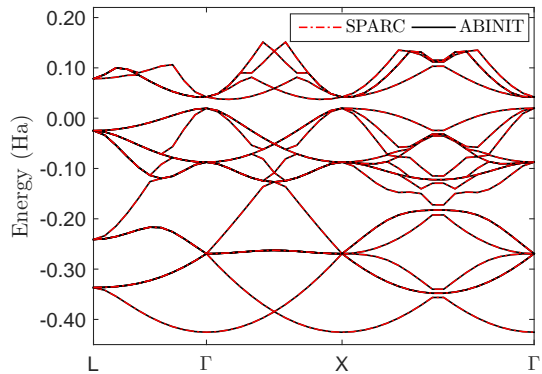}\\
\caption{Band structure plot for bulk silicon.}
\label{fig:BandStructure}
\end{figure}

Next, we compare the band structure plot at the equilibrium lattice constants determined above, i.e., $10.157$ Bohr for SPARC and $10.160$ Bohr for ABINIT. Specifically, we choose the $L-\Gamma-X-\Gamma$ circuit, whose coordinates in terms of the reciprocal lattice vectors are [$-0.5$ $0.5$ $0.5$], [$0.0$ $0.0$ $0.0$], [$1.0$ $0.0$ $0.0$], and [$1.0$ $1.0$ $1.0$], respectively. We discretize the $L-\Gamma$, $\Gamma-X$, and $X-\Gamma$ line segments into $10$, $12$, and $17$ divisions, respectively. At each resulting $\bk$-point, we determine the band structure (at the electronic ground-state) in SPARC by repeating the CheFSI steps until convergence. In Fig. \ref{fig:BandStructure}, we present the band structure plot so computed by SPARC and compare it with that calculated by ABINIT. It is clear that there is very good agreement, with the curves being nearly identical. In particular, the HOMO eigenvalue, LUMO eigenvalue, and bandgap are in agreement to within $7 \times 10^{-5}$ Ha, $1 \times 10^{-5}$ Ha, and $6 \times 10^{-5}$ Ha, respectively. 

\subsection{Slab and wire properties} \label{Subsec:SlabWireProp}
Here, we verify ability of SPARC to accurately calculate the properties of slabs and wires. For this purpose, we choose the aluminum slab ($1 \times 1 \times 1$ Brillouin zone integration) and silicon wire ($4 \times 1 \times 1$ Brillouin zone integration) described in Section \ref{Subsection:Convergence:Vacuum} as representative examples. In SPARC, we utilize a vacuum of $6.95$ Bohr and $5.59$ Bohr for the slab and wire, respectively. In ABINIT, the corresponding values are $12.55$ Bohr and $10$ Bohr, respectively. The energy and forces so computed in ABINIT are accurate to within $1 \times 10^{-4}$ Ha/atom and $1 \times 10^{-4}$ Ha/Bohr, respectively.\footnote{As shown in Section \ref{Subsection:Convergence:Vacuum}, there is stagnation in the convergence of the energy and atomic forces with vacuum size in ABINIT, which limits its accuracy.} 

\begin{figure}[H]
\centering
\subfloat[Aluminum slab: $y=-3.89$ Bohr plane]{\label{fig:AluminumSlabEdge}\includegraphics[trim =3cm 0cm 3.5cm 1cm,clip,keepaspectratio=true,width=0.35\textwidth]{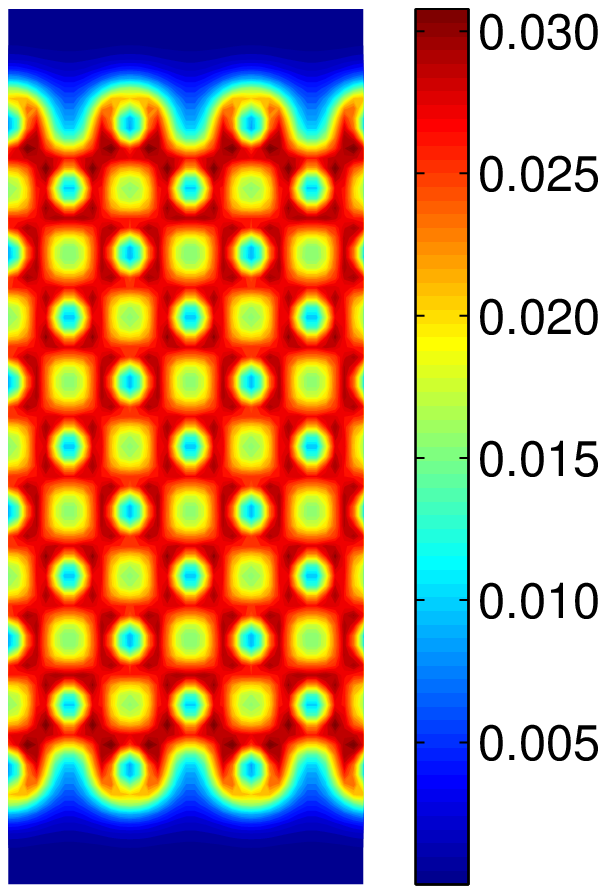}} 
\subfloat[Silicon wire: $x=0$ Bohr plane]{\label{fig:SiWireTransverse}\includegraphics[trim =0cm 0cm 0cm 0cm,clip,keepaspectratio=true,width=0.48\textwidth]{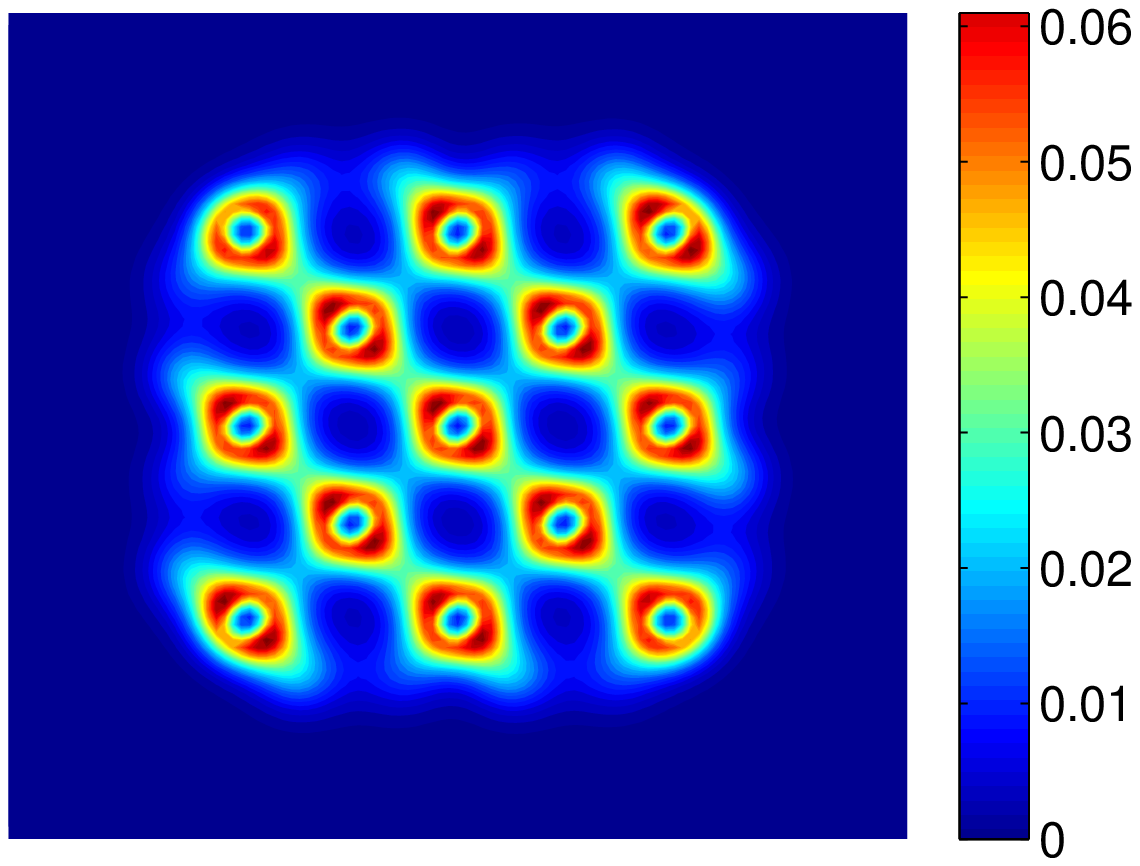}}
\caption{Electron density contours for the aluminum slab and silicon wire.}
\label{fig:Contours:SlabWire}
\end{figure}

First, we calculate the unrelaxed surface energy of the aluminum slab:
\begin{equation}
\gamma = N \left( \frac{\hat{\mathcal{F}}_{slab} - \hat{\mathcal{F}}_{bulk}}{2A} \right) \,,
\end{equation} 
where $\hat{\mathcal{F}}$ is the free energy/atom of the slab, and $\hat{\mathcal{F}}_{bulk}$ is the free energy/atom of bulk aluminum. The surface energies so calculated by SPARC and ABINIT are in agreement to within $0.001$ J/m$^2$. In addition, the maximum difference in the atomic forces is within $3 \times 10^{-4}$ Ha/Bohr. In Fig. \ref{fig:AluminumSlabEdge}, we present the contours of the electron density computed by SPARC on the $y=-3.89$ Bohr plane of the slab. 

\begin{figure}[H]
\centering
\includegraphics[trim =0cm 0.5cm 0cm 0cm,clip,keepaspectratio=true,width=0.45\textwidth]{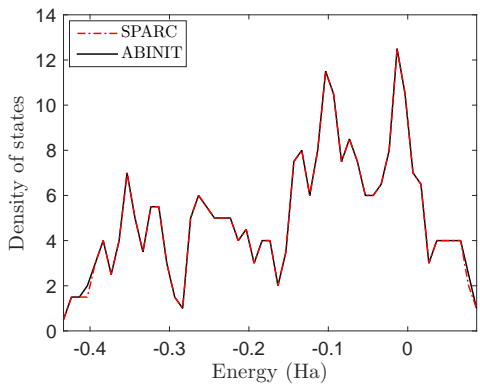}\\
\caption{Density of states for the silicon wire.}
\label{fig:DOS}
\end{figure}

Next, we determine the density of states (DOS) of the silicon wire using the relation
\begin{equation}
 D(\mathcal{E}) = 2 \sum_{n=1}^{N_s} \fint_{BZ} \delta(\lambda_n(\bk)-\mathcal{E}) \, \mathrm{d\bk} \,\,,
\end{equation}
where $\delta$ is the Dirac delta function. In Fig. \ref{fig:DOS}, we present the DOS computed by SPARC and ABINIT. It is clear that there is excellent agreement between SPARC and ABINIT, with the curves being practically indistinguishable. In addition, the energy and atomic forces are in agreement to within $7 \times 10^{-5}$ Ha/atom and $3 \times 10^{-4}$ Ha/Bohr, respectively. In Fig. \ref{fig:SiWireTransverse}, we present the contours of the electron density computed by SPARC on the $x=0$ Bohr plane of the wire.

\subsection{Geometry optimization} \label{Subsec:GeometryOptimization}
\begin{figure}[H]
\centering
\subfloat[Computed energy and its cubic spline fit]{\label{fig:EnergyDisplacement}\includegraphics[keepaspectratio=true,width=0.48\textwidth]{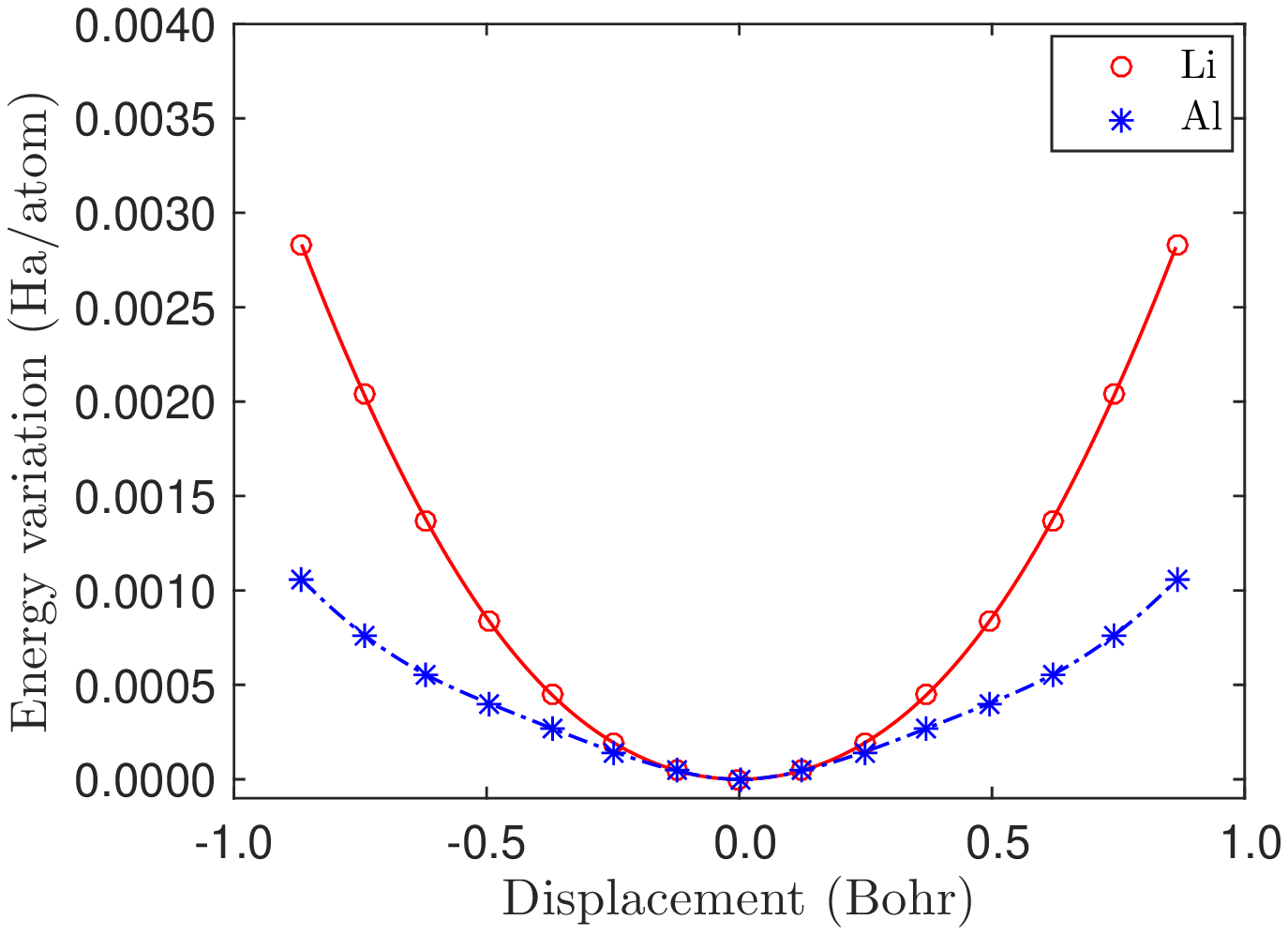}}
\subfloat[Computed force and the derivative of the cubic spline fit to the energy]{\label{fig:ConsistencyEnergyForce}\includegraphics[keepaspectratio=true,width=0.48\textwidth]{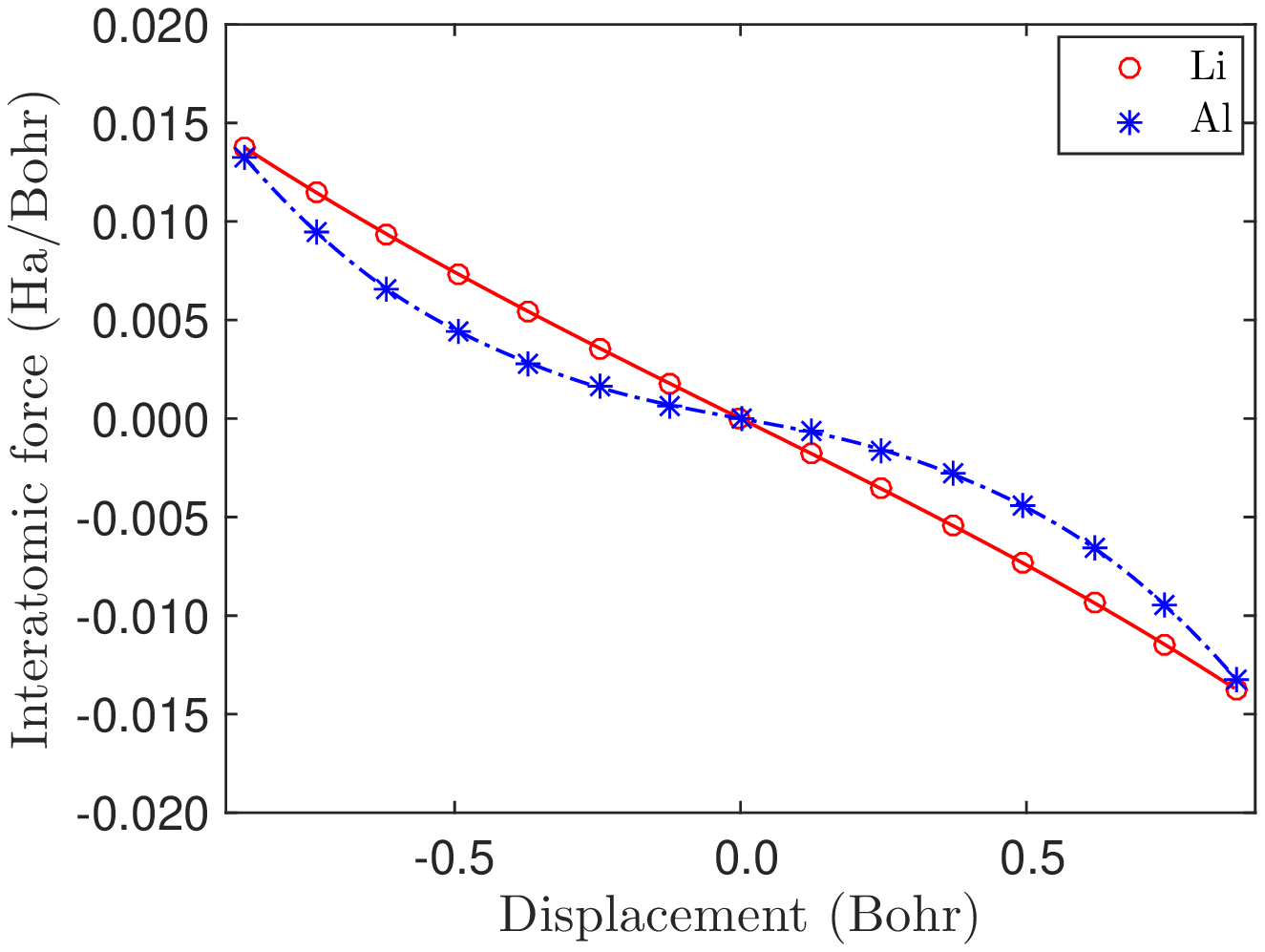}}
\caption{Variation in the energy and atomic force as a function of atomic displacement. In lithium, the body centered lithium atom is displaced along the body diagonal. In aluminum, the corner atom is displaced along the body diagonal.}
\label{Fig:EnergyForceConsistency}
\end{figure}

We now verify the capacity of SPARC to perform accurate geometry optimizations. To do so, we first check the consistency of the atomic forces with the energy. As representative examples, we select single unit cells of lithium and aluminum with lattice constants of $6.59$ Bohr and $8$ Bohr, respectively, and discretize them using mesh sizes of $h=0.549$ Bohr and $0.615$ Bohr, respectively. In Fig. \ref{Fig:EnergyForceConsistency}, we plot the variation in energy and force when the body centered lithium atom is displaced along the body diagonal, and the corner aluminum atom is displaced along the body diagonal. Specifically, in Fig. \ref{fig:EnergyDisplacement}, we plot the computed energy and its curve fit using cubic splines. In Fig. \ref{fig:ConsistencyEnergyForce}, we plot the computed atomic force and the derivative of the cubic spline fit to the energy. The evident agreement demonstrates that the computed energy and atomic forces are indeed consistent. Moreover, there is no noticeable `egg-box' effect \cite{brazdova2013atomistic}---a phenomenon arising due to the breaking of the translational symmetry---at meshes required for obtaining chemical accuracies.\footnote{This effect can further be reduced by using the double grid method \cite{OnoHir99}, high-order spatial integration \cite{BobSchChe15}, or suitably modifying the pseudopotential \cite{OCTOPUS,briggs1996real}}

\begin{figure}[H]
\centering
\subfloat[$3 \times 3 \times 3$ unit cells of lithium with a vacancy]{\label{fig:ElecDensityLi}\includegraphics[trim =0cm 1.6cm 0cm 0cm,clip,keepaspectratio=true,width=0.48\textwidth]{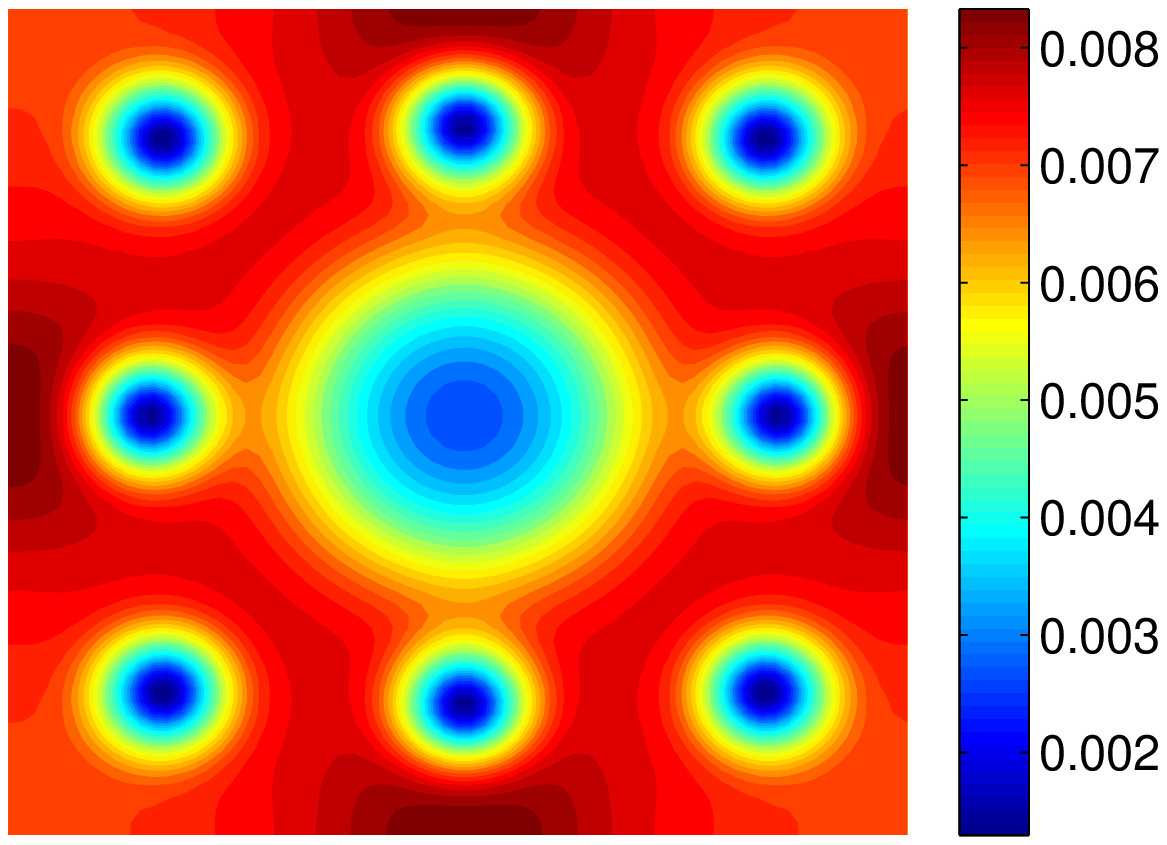}}
\subfloat[$2 \times 2 \times 2$ unit cells of aluminum with a vacancy]{\label{fig:ElecDensityAl}\includegraphics[trim =0cm 1.6cm 0cm 0cm,clip,keepaspectratio=true,width=0.48\textwidth]{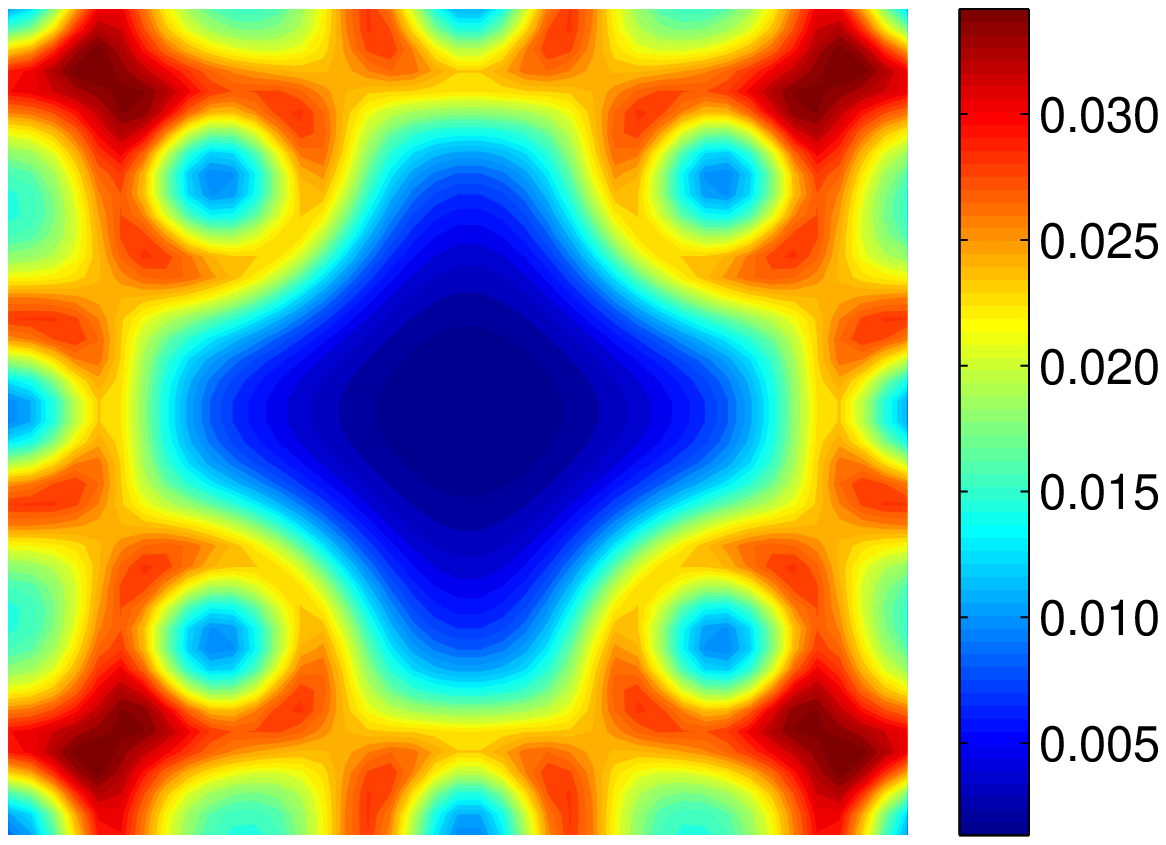}}
\caption{Mid-plane electron density contours.}
\label{Fig:RhoContour}
\end{figure}

Next, we determine the overall ground-state for two systems: $3 \times 3 \times 3$ unit cells of lithium and $2 \times 2 \times 2$ unit cells of aluminum, both with a vacancy. We compare the results with ABINIT, wherein we use a plane wave cutoff of $30$ Ha, which results in energy and forces that are converged to within $5 \times 10^{-6}$ Ha/atom and $5 \times 10^{-6}$ Ha/Bohr, respectively. We calculate the vacancy formation energy $\mathcal{E}_{vf}$ using the relation \cite{gillan1989calculation}
\begin{equation}
\mathcal{E}_{vf} =\mathcal{F}_0\left(N-1,1,\frac{N-1}{N} \Omega \right) - \left( \frac{N-1}{N} \right)\mathcal{F}_0(N,0,\Omega) \,,
\end{equation}
where $\mathcal{F}_0(N,n_v,\Omega)$ denotes the energy of the system with $N$ occupied lattice sites and $n_v$ vacancies. The vacancy formation energy so computed by SPARC and ABINIT is in agreement to within $7 \times 10^{-4}$ Ha. In addition, the fully relaxed atomic positions differ by no more than $1.6 \times 10^{-3}$ Bohr. On refining the mesh, the agreement between SPARC and ABINIT further improves. For example, consider a mesh-size of $h=0.284$ Bohr and $h=0.390$ Bohr for lithium and aluminum, respectively. The vacancy formation energy so computed by SPARC and ABINIT are in agreement to within $9 \times 10^{-6}$ Ha, and the maximum difference in the final atomic positions is $1.3 \times 10^{-4}$ Bohr. The contours of electron density on the mid-plane of these systems are plotted in Fig. \ref{Fig:RhoContour}. 

\subsection{Molecular dynamics}
We now study the capability of SPARC to perform accurate Born-Oppenheimer molecular dynamics calculations \cite{CPMD}. Specifically, we perform a $1$ ps NVE molecular dynamics simulation for the aluminum system consisting of $5 \times 5 \times 5$ FCC unit cells ($500$ atoms). We utilize a mesh-size of $h=0.667$ Bohr, an initial ionic temperature of $T = 3157.75$ K, and a time step of $1$ fs. We assign identical initial velocities to all the atoms and set the initial accelerations to be zero. At every molecular dynamics step, we set the electronic temperature (smearing) to be equal to the ionic temperature. 

\begin{figure}[H]
\centering
\subfloat[Total energy]{\label{Fig:MD:TE} \includegraphics[keepaspectratio=true,width=0.48\textwidth]{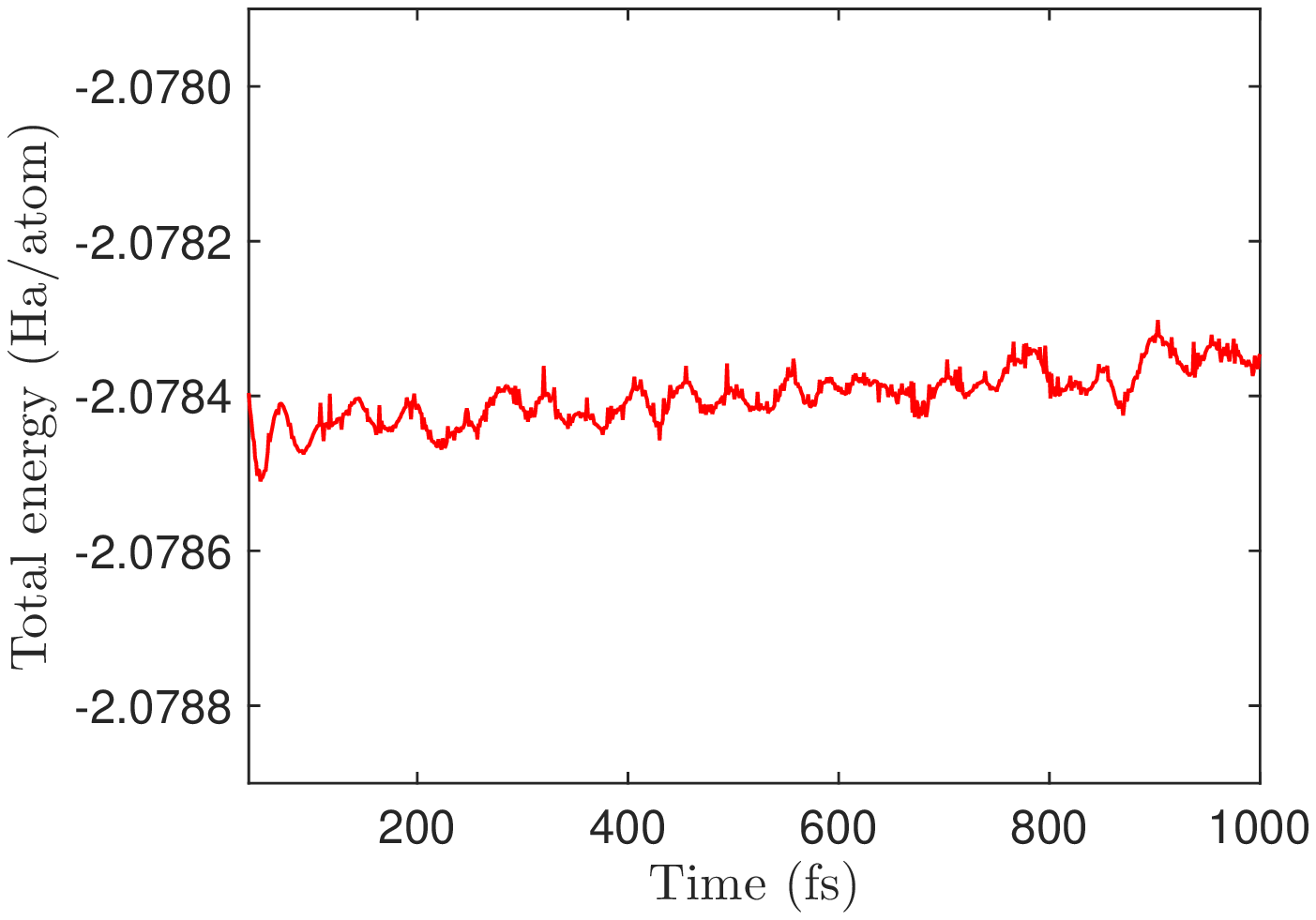}} 
\subfloat[Free energy]{\label{Fig:MD:FE} \includegraphics[keepaspectratio=true,width=0.48\textwidth]{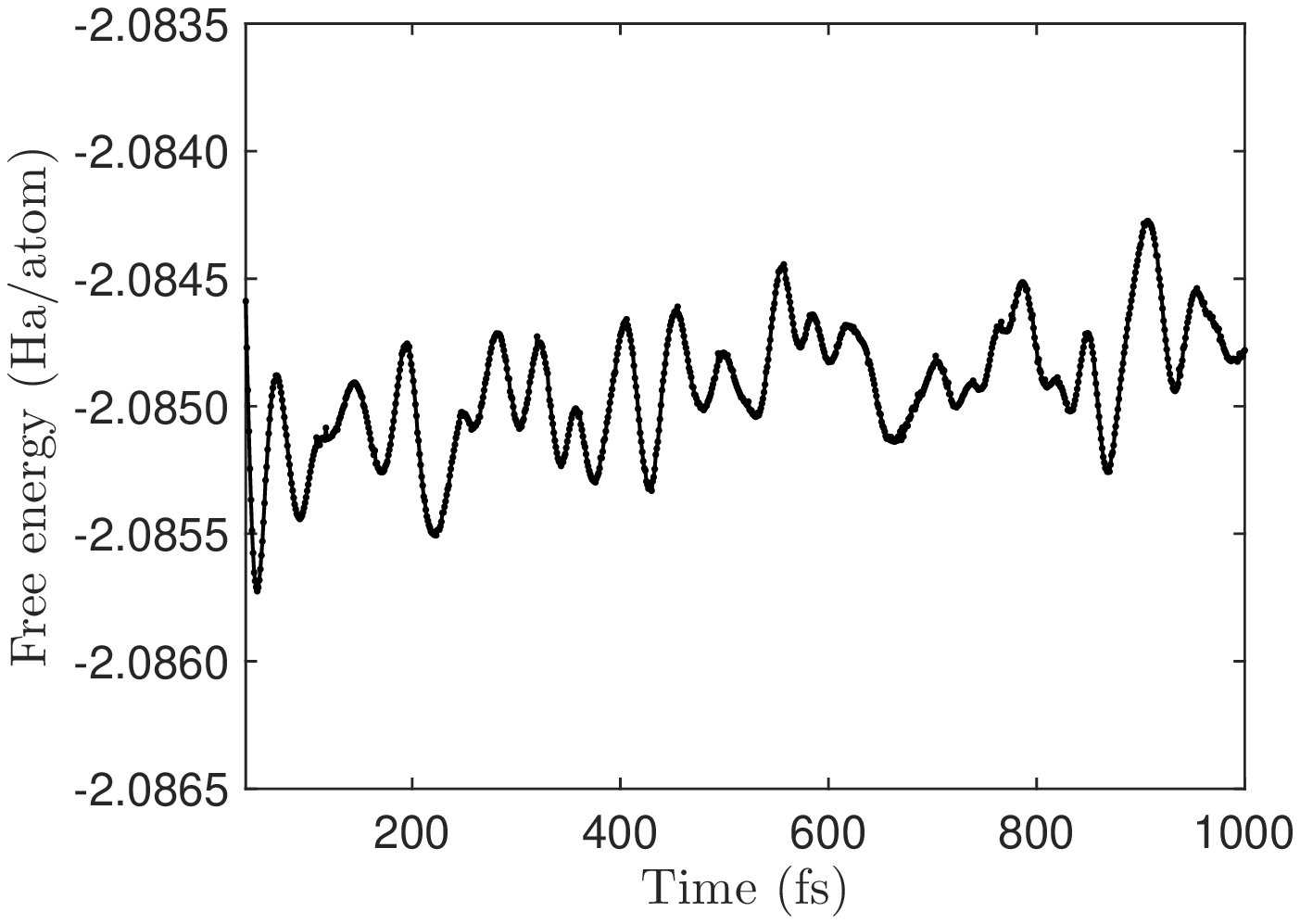}} \\
\subfloat[Kinetic energy]{\label{Fig:MD:KE} \includegraphics[keepaspectratio=true,width=0.48\textwidth]{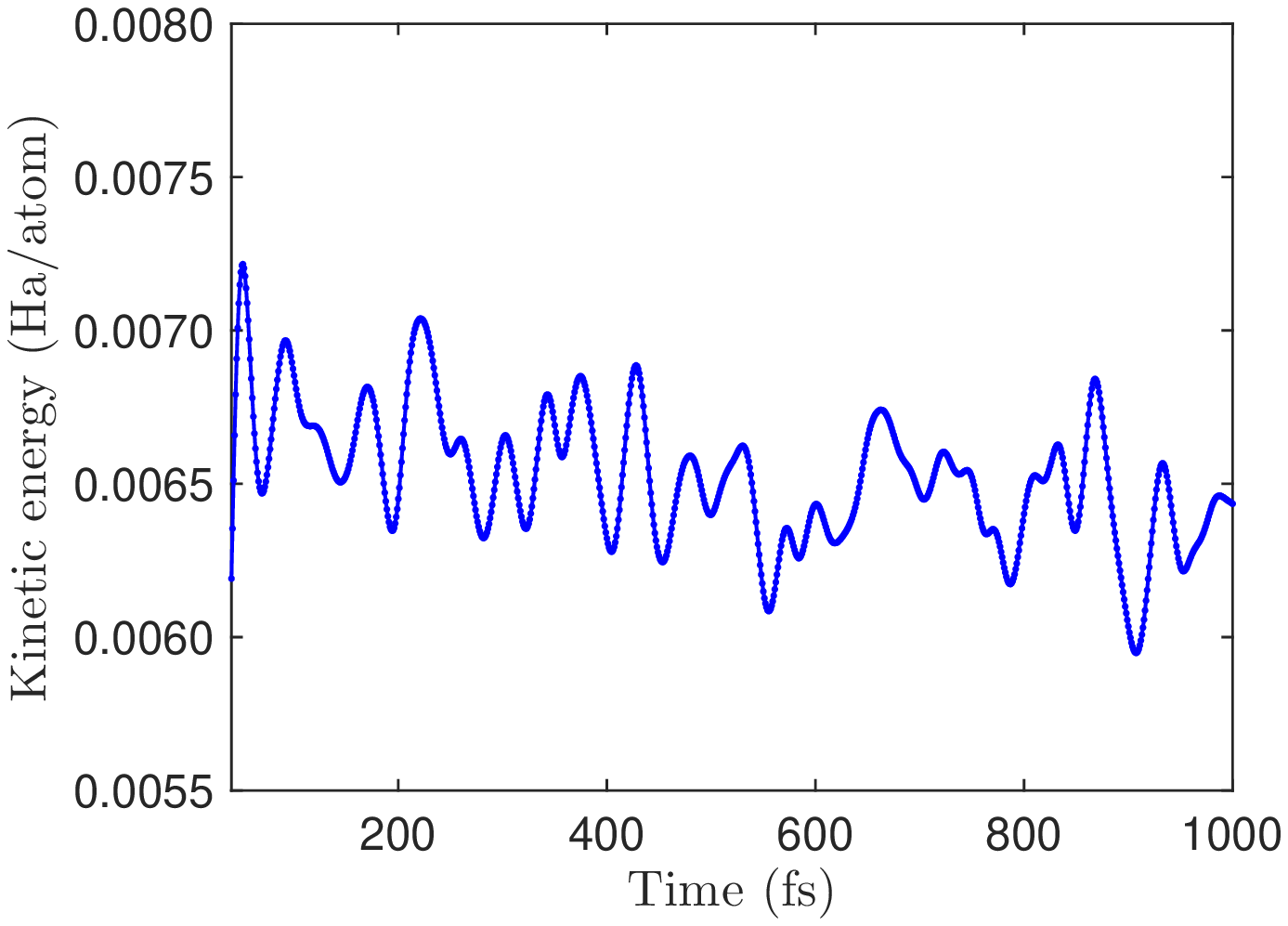}} 
\subfloat[Temperature]{\label{Fig:MD:Temperature} \includegraphics[keepaspectratio=true,width=0.48\textwidth]{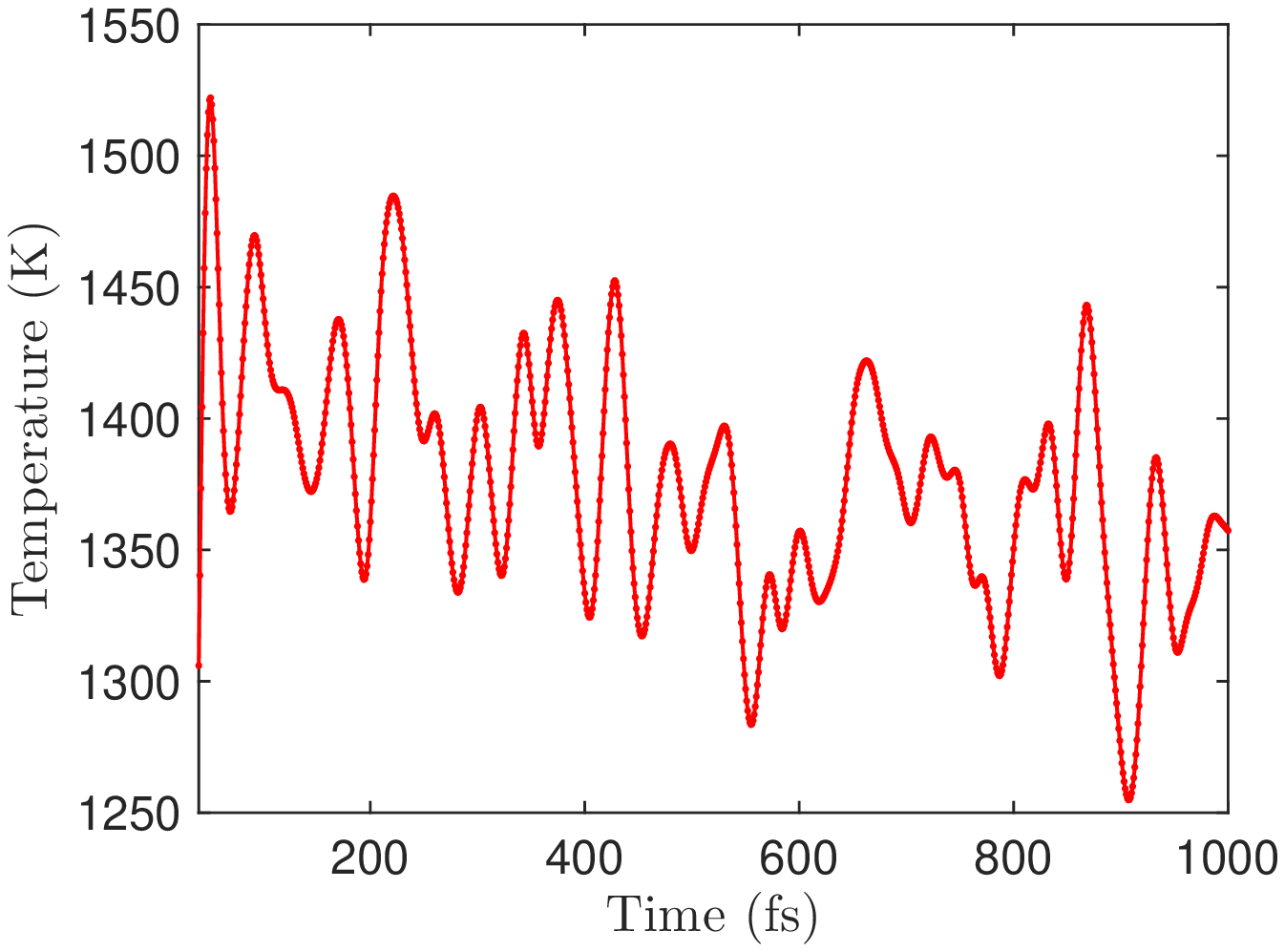}} 
\caption{The variation of total energy, free energy, kinetic energy and temperature during the NVE molecular dynamics simulation for $5 \times 5 \times 5$ FCC unit cells of alumimum at an initial temperature of $T = 3157.75$ K.}
\label{Fig:MolecularDynamics}
\end{figure}

In Fig. \ref{Fig:MolecularDynamics}, we plot the variation in the energy (free energy, kinetic energy, and total energy) and temperature as a function of time. The plots start at $40$ fs, which corresponds to the time required to achieve statistical equilibrium. During the simulation, the mean and standard deviation of the total energy is $-2.07842$ Ha/atom and $1.6 \times 10^{-4}$ Ha/atom, respectively. The drift in total energy as obtained from the linear fit to the data is $1 \times 10^{-4}$ Ha/atom-ps.\footnote{We have found that the drift in the total energy further reduces as the spatial discretization is refined.} Overall, the lack of any significant drift verifies that there are no systematic errors in SPARC, which further confirms the accuracy of the energy and atomic forces in SPARC. 

\subsection{Scaling and Performance} \label{Subsec:ScalingAndPerformance} 
Having verified the accuracy of SPARC relative to ABINIT in previous subsections, we now compare their efficiency\footnote{A comparison of the accuracy and efficiency of SPARC with other finite-difference DFT codes is presented in Appendix \ref{Appendix:RealSpaceComparison}.}. As representative systems, we choose $n \times n \times n$ ($n \in \{3,4,5,6,7\}$) unit cells of aluminum with a vacancy. In SPARC, we employ a mesh-size of $h=0.778$ Bohr and Chebyshev polynomial filter of degree $15$. In ABINIT, we use a plane-wave energy cutoff of $9$ Ha. We choose all the other parameters in both codes so as to achieve an overall accuracy of $5 \times 10^{-4}$ Ha/atom and $5 \times 10^{-4}$ Ha/Bohr in the energy and forces, respectively. The times reported here include the calculation of the electronic ground-state as well as the atomic forces, i.e., geometry optimization is not performed. 

\begin{figure}[H]
\centering
\subfloat[Strong scaling]{\label{Fig:StrongScaling}\includegraphics[keepaspectratio=true,width=0.48\textwidth]{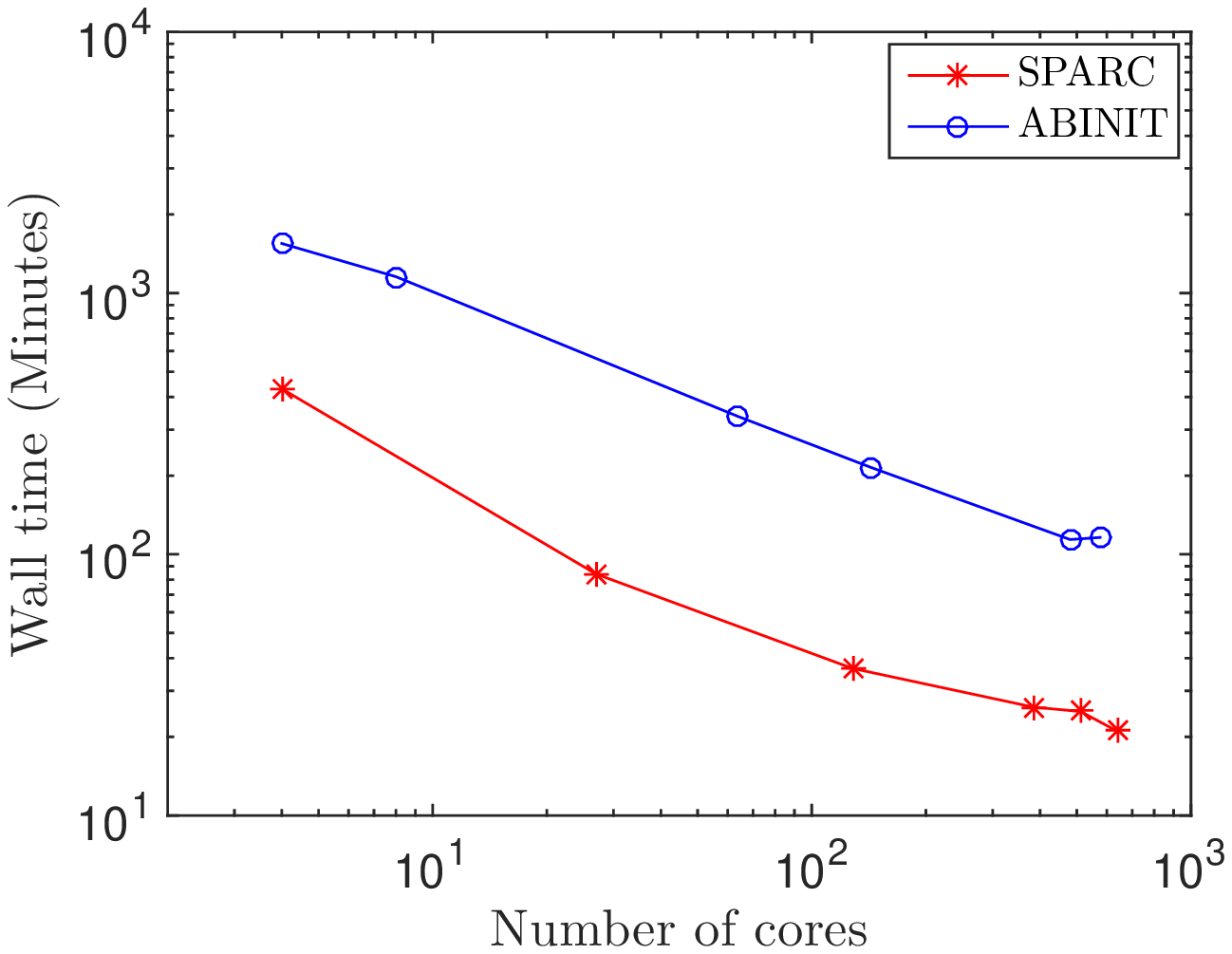}}
\subfloat[Weak scaling]{\label{Fig:WeakScaling}\includegraphics[keepaspectratio=true,width=0.48\textwidth]{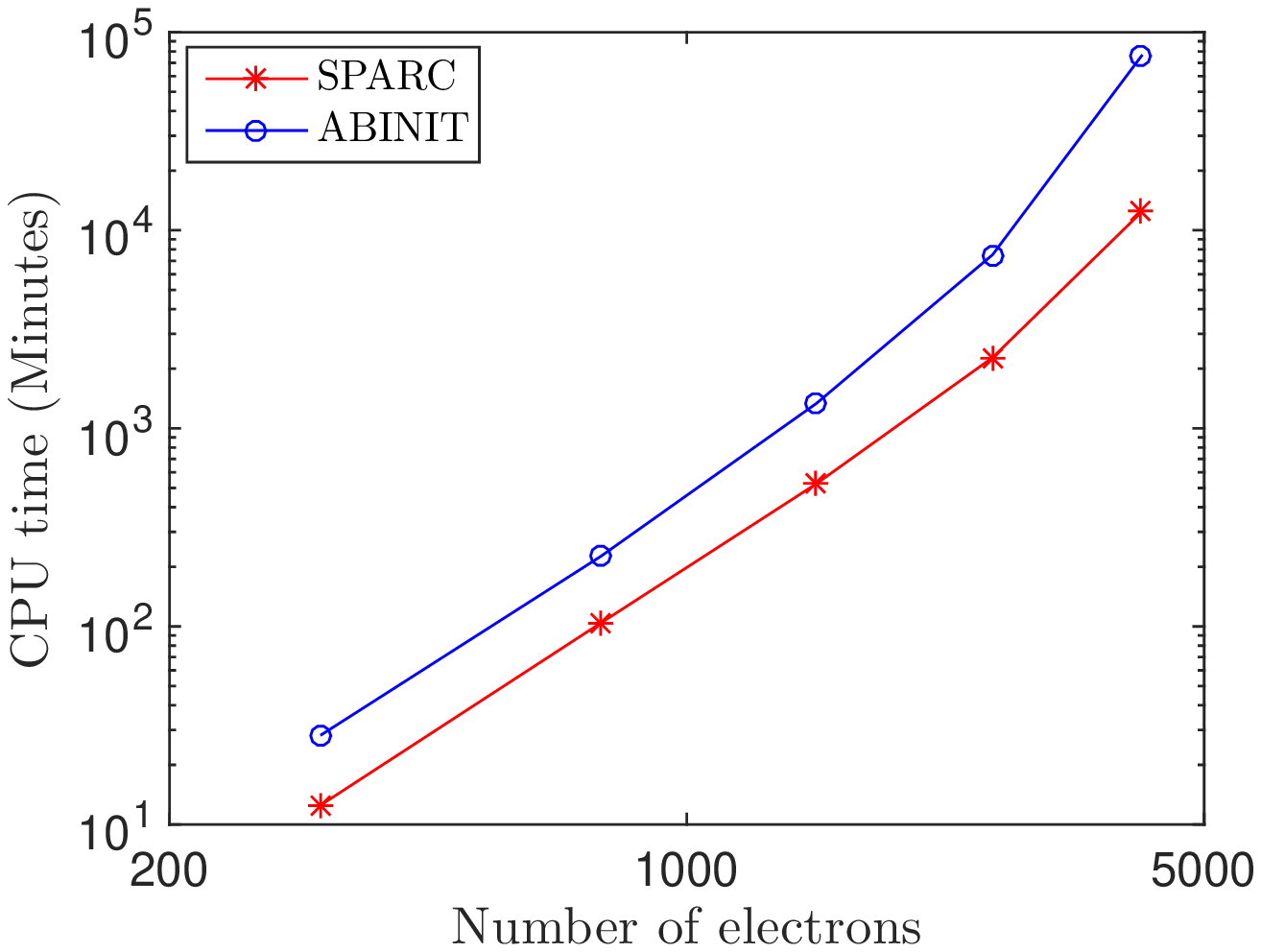}}
\caption{Strong and weak scaling for SPARC and ABINIT. The system utilized for strong scaling is $6\times 6\times 6$ FCC unit cells of aluminum with a vacancy. The systems employed for weak scaling are $3\times 3\times 3$, $4\times 4\times 4$, $5\times 5\times 5$, $6\times 6\times 6$, and $7\times 7\times 7$ unit cells of aluminum, each with a vacancy.}
\end{figure}

First, we compare the strong scaling of SPARC and ABINIT for $6\times 6\times 6$ FCC unit cells of aluminum with a vacancy. We utilize $4$, $8$, $64$, $144$, $480$, and $576$ cores for performing the simulation with ABINIT, which it suggests is optimal in the range of $1$ to $1000$ cores. For SPARC, we select $4$, $8$, $27$, $128$, $384$, and $512$ cores. In Fig. \ref{Fig:StrongScaling}, we plot the wall time taken by SPARC and ABINIT as a function of the number of processors.  We observe that both SPARC and ABINIT display similar trends with respect to strong scaling, with curves being close to parallel and no further reduction in wall time observed after approximately $600$ cores. However, the prefactors of SPARC are significantly smaller, by up to a factor of $4$.

Next, we compare the weak scaling of SPARC with ABINIT for $3\times 3\times 3$, $4\times 4\times 4$, $5\times 5\times 5$, $6\times 6\times 6$, and $7\times 7\times 7$ unit cells of aluminum, each with a vacancy. The number of electrons in these systems range from $321$ to $4116$. For both SPARC and ABINIT, we fix the number of electrons per core to be approximately $96$, and choose at most $4$ cores from every compute node.  In Figure \ref{Fig:WeakScaling}, we present the results so obtained for the variation in total CPU time versus the number of electrons. We determine the scaling for SPARC and ABINIT to be $\mathcal{O}(N_e^{2.64})$ and $\mathcal{O}(N_e^{2.99})$ respectively. The prefactor for SPARC is again noticeably smaller, with speedups over ABINIT ranging from factors of $2.2$ to $6$. 

Finally, we compare the minimum wall time achievable by SPARC and ABINIT for the aforementioned systems. We restrict the maximum number of electrons per core to $96$. In SPARC, we choose the number of cores as multiples of $32$, whereas we select the number of cores and parallelization scheme in ABINIT that it suggests are optimal. In Table \ref{Table:Time}, we present the results for the minimum wall time so achieved. We observe that SPARC outperforms ABINIT by factors larger than $2.5$  for all the systems considered. In particular, SPARC requires a factor of approximately $5$ less wall time for the largest system. Overall, we conclude that SPARC is a very efficient DFT formulation and implementation that is very competitive with currently existing highly optimized plane-wave codes for extended systems.\footnote{In previous work \cite{sparc2016cluster}, we have already shown that SPARC is highly competitive with plane-wave codes for isolated systems.}

\begin{table}[H]
\centering
\begin{tabular}{ccc}
\hline
System &   SPARC     & ABINIT  \\   
\hline
$3 \times 3 \times 3$    &   $0.43$ $(64)$    & $1.77$ $(188)$   \\
$4 \times 4 \times 4$    &   $1.69$ $(96)$   & $5.67$ $(320)$          \\
$5 \times 5 \times 5$   &   $6.99$ $(256)$  & $17.42$ $(396)$   \\
$6 \times 6 \times 6$    &   $20.98$ $(640)$ & $113.87$ $(480)$ \\
$7 \times 7 \times 7$   &   $79.15$ $(704)$ & $398.77$ $(795)$ \\
\hline
\end{tabular}
\caption{Minimum wall time in minutes for $n\times n \times n$ ($n \in \{3,4,5,6,7\}$) FCC unit cells of aluminum with a vacancy. The number in brackets denotes the number of cores on which the minimum wall time is achieved.}
\label{Table:Time}
\end{table} 

\section{Concluding Remarks} \label{Sec:Conclusions} 
In this work, we have extended the capabilities of SPARC (Simulation Package for Ab-initio Real-space Calculations)---finite-difference formulation and parallel implementation of Density Functional Theory (DFT)---to enable the study of extended systems like crystals, slabs, and wires. Specifically, utilizing the Chebyshev polynomial filtered self-consistent field iteration in combination with the reformulation of the electrostatics and the non-local atomic force component, we have developed a framework that enables the efficient evaluation of energies and atomic forces to within the desired accuracies in DFT while employing the finite-difference representation. 

Using a wide variety of materials systems, we have demonstrated that SPARC obtains systematic and high rates of convergence in the energy and forces with mesh-size to reference plane-wave results; exponential convergence with vacuum size for slabs and wires; energy and forces that are consistent and have negligible `egg-box' effect; accurate ground-state properties; and negligible drift in molecular dynamics simulations. We have also shown that SPARC displays weak and strong scaling that is similar to well-established and optimized plane-wave codes for systems having up to thousands of electrons, but with a noticeably smaller prefactor. Additionally, we have established that SPARC significantly outperforms other finite-difference DFT packages.

We conclude by noting that there is scope for significant improvement of the current SPARC implementation.  In particular, enabling the parallelization over bands is expected to enhance the performance of SPARC, especially in the context of strong scaling. Further, incorporating scalable diagonalization techniques for the solution of the subspace eigenproblem in parallel will enable the study of significantly larger systems. These improvements along with additional optimization of code are expected to further improve the efficiency of SPARC. Finally, removing the dependency on external packages (i.e., PETSc and MKL) is expected to significantly improve the portability and usability of SPARC, making it another worthy subject for future work.


\section*{Acknowledgements}
\noindent The authors gratefully acknowledge the support of National Science Foundation under Grant Number $1333500$. The authors also gratefully acknowledge the valuable comments and suggestions of the anonymous referee.


\bibliographystyle{ReferenceStyle}

\vspace{15mm}

\appendix

{\LARGE \bf Appendix}


\section{Electrostatic correction for overlapping pseudocharge densities} \label{Appendix:Correct:RepulsiveEnergy}
In ab-initio calculations, even when the pseudopotential approximation is employed, the repulsive energy is still calculated by treating the nuclei as point charges. Since the electrostatic formulation in this work does not make this distinction, there is disagreement with convention when the pseudocharge densities overlap. The correction which reestablishes agreement can be written as \cite{Suryanarayana2014524}
\begin{eqnarray} 
E_c(\bR) & = & \frac{1}{2} \int_{\Omega} \left( \tilde{b}(\bx,\bR) + b(\bx,\bR) \right) V_c(\bx,\bR) \, \mathrm{d\bx} + \frac{1}{2}\sum_{I} \int_{\Omega} b_I(\bx,\bR_I) V_I(\bx,\bR_I) \, \mathrm{d\bx} \nonumber \\
& & - \frac{1}{2}\sum_{I} \int_{\Omega} \tilde{b}_I(\bx,\bR_I) \tilde{V}_I(\bx,\bR_I) \, \mathrm{d\bx} \,, \label{Eqn:RepulsiveCorrectionEnergy}
\end{eqnarray} 
where the summation index $I$ runs over all atoms in $\R^3$, and 
\begin{equation} \label{Eqn:Vc}
 V_c(\bx,\bR) =  \sum_{I}\left(\tilde{V}_I(\bx,\bR_I)-V_I(\bx,\bR_I)\right) \,.
\end{equation} 
In addition, $\tilde{b}$ denotes the reference pseudocharge density, and $\tilde{b}_I$ represents the spherically symmetric and compactly supported reference charge density of the $I^{th}$ nucleus that generates the potential $\tilde{V}_I$, i.e., 
\begin{eqnarray} 
\tilde{b}(\bx,\bR) = \sum_{I} \tilde{b}_I(\bx,\bR_J) \,, & & \tilde{b}_I(\bx,\bR_I) = - \frac{1}{4 \pi} \nabla^2 \tilde{V}_I(\bx,\bR_I) \,, \label{Eqn:RefPseudochargeDefinition} \\
\int_{\Omega} \tilde{b}(\bx,\bR) \, \mathrm{d\bx} = - N_e \,, & &  \int_{\R^3} \tilde{b}_I(\bx,\bR_I) \, \mathrm{d\bx} = Z_I  \,. \label{Eqn:IntegralPseudochargesRef}
\end{eqnarray}
The discrete form of the repulsive energy correction in Eqn. \ref{Eqn:RepulsiveCorrectionEnergy} takes the form 
\begin{equation}
E_c^h = \frac{1}{2} h_1h_2h_3 \sum_{i=1}^{n_1} \sum_{j=1}^{n_2} \sum_{k=1}^{n_3} \left( (\tilde{b}^{(i,j,k)} + b^{(i,j,k)}) V_c^{(i,j,k)} + \sum_{I} b_I^{(i,j,k)} V_I^{(i,j,k)} - \sum_{I} \tilde{b}_I^{(i,j,k)} \tilde{V}_{I}^{(i,j,k)}  \right) \,, \label{Eqn:Ec:Disc}
\end{equation}
where the integrals have been approximated using the trapezoidal rule in Eqn. \ref{Eqn:IntApprox}. 

The correction to the atomic forces can be written as \cite{Suryanarayana2014524}
\begin{eqnarray}
\mathbf{f}_{J,c}(\bR) & = & \frac{1}{2} \sum_{J'} \int_{\Omega} \bigg[ \nabla \tilde{b}_{J'}(\bx,\bR_{J'}) \left(V_c(\bx,\bR)- \tilde{V}_{J'}(\bx,\bR_{J'})\right) + \nabla b_{J'}(\bx,\bR_{J'}) \left(V_c(\bx,\bR)+V_{J'}(\bx,\bR_{J'})\right) \nonumber \\ 
& + & \left(\nabla \tilde{V}_{J'}(\bx,\bR_{J'}) - \nabla V_{J'}(\bx,\bR_{J'})\right) \left(\tilde{b}(\bx,\bR)+b(\bx,\bR)\right) + b_{J'}(\bx,\bR_{J'}) \nabla V_{J'}(\bx,\bR_{J'}) \label{Eqn:RepulsiveCorrectionForce} \\ 
& - & \tilde{b}_{J'}(\bx,\bR_{J'}) \nabla \tilde{V}_{J'}(\bx,\bR_{J'}) \bigg] \,\mathrm{d\bx} \,,  \nonumber
\end{eqnarray}
whose discrete form: 
\begin{eqnarray}
\mathbf{f}_{J,c}^h & = & \frac{1}{2} h_1h_2h_3 \sum_{J'} \sum_{i=1}^{n_1} \sum_{j=1}^{n_2} \sum_{k=1}^{n_3} \bigg( \nabla_h \tilde{b}_{J'}\big|^{(i,j,k)} \left(V_{c}^{(i,j,k)} - \tilde{V}_{J'}^{(i,j,k)}\right) +  \nabla_h b_{J'}\big|^{(i,j,k)} \left(V_{c}^{(i,j,k)}+  V_{J'}^{(i,j,k)}\right)  \nonumber \\ 
& + & \nabla_h (\tilde{V}_{J'}^{(i,j,k)} - V_{J'}^{(i,j,k)}) \big|^{(i,j,k)} \left(\tilde{b}^{(i,j,k)}+b^{(i,j,k)}\right) + b_{J'}^{(i,j,k)} \nabla_h V_{J'}\big|^{(i,j,k)} - \tilde{b}_{J'}^{(i,j,k)} \nabla_h \tilde{V}_{J'}\big|^{(i,j,k)}  \bigg) \,. \label{Eqn:LF:Disc} \nonumber \\
\end{eqnarray} 
As before, the summation $J'$ runs over the $J^{th}$ atom and its periodic images. For the reference potential, we choose the potential that has been previously employed for generating neutralizing densities in all-electron calculations \cite{Pask2012}.

\section{Pseudopotential parameters} \label{Appendix:Pseudopotential}
The cutoff radii $r_J^c$ employed in this work for the different angular momentum components of the Troullier-Martins pseudopotential are listed in Table \ref{Table:AluminumClustersForce}. The $l=0$ component is chosen to be local in all the calculations. 

\begin{table}[H]
\centering
\begin{tabular}{ccccc}
\hline 

Atom type & \multicolumn{3}{c}{Radial cutoff (Bohr)}  \\
          &  $l=0$        & $l=1$        & $l=2$         \\
\hline
H         & $1.25$         &  $-$        & $-$               \\
Li         & $2.40$         &  $2.40$        & $-$               \\
Al        & $2.60$         &  $2.60$        & $-$              \\
Si        & $1.80$         &  $1.80$        & $1.80$            \\
Au         & $2.60$         &  $2.60$        & $2.60$            \\
\hline
\end{tabular}
\caption{Cutoff radii for non-local projectors within the Troullier-Martins pseudopotential.}
\label{Table:AluminumClustersForce}
\end{table}


\section{Discrete pseudocharge density properties} \label{Appendix:Pseudocharge}
\begin{figure}[H]
\centering
\includegraphics[keepaspectratio=true,width=0.46\textwidth]{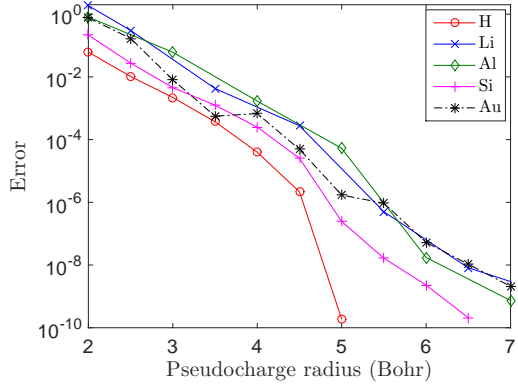}
\caption{Normalized error in the net enclosed charge as a function of pseudocharge radius for $h=0.5$ Bohr.}
\label{Fig:PseudochargeDecay} 
\end{figure}

The continuous pseudocharge density of the $J^{th}$ atom has compact support in a sphere of radius $r_J^c$---cutoff radius for the local pseudopotential---centered at that atom. However, the corresponding discrete pseudocharge density has infinite extent due to the use of the finite-difference approximated Laplacian. In Fig. \ref{Fig:PseudochargeDecay}, for a mesh-size $h=0.5$ Bohr, we plot the normalized error in the net enclosed charge as a function of the pseudocharge radius $r_J^b$. It is clear that there is exponential decay, which allows for truncation at some finite radius without significant loss of accuracy.

\begin{figure}[H]
\centering
\includegraphics[keepaspectratio=true,width=0.46\textwidth]{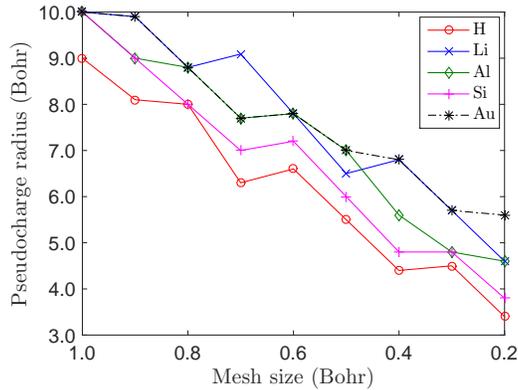}
\caption{Variation of pseudocharge radius as a function of mesh size.}
\label{Fig:PseudochargeRadii} 
\end{figure}

In Fig. \ref{Fig:PseudochargeRadii}, we plot the truncation radius $r_J^b$ required to achieve the accuracy of $\varepsilon_b= 10^{-8}$ (Eqn. \ref{Eqn:PseudochargeRadiusChoice}) as a function of mesh-size $h$. We observe that $r_J^b$ becomes smaller as the discretization is refined, with $r_J^b \rightarrow r_J^c$ as $h \rightarrow 0$. The slight non-monotonicity observed at some places in the curves arises due to the fact that $r_J^b$ is chosen to be a multiple of $h$ within SPARC.  

\section{Comparison of SPARC with other finite-difference DFT codes} \label{Appendix:RealSpaceComparison}
\begin{figure}[H]
\centering
\subfloat[Energy]{\label{fig:energyConvergence_rs}\includegraphics[keepaspectratio=true,width=0.46\textwidth]{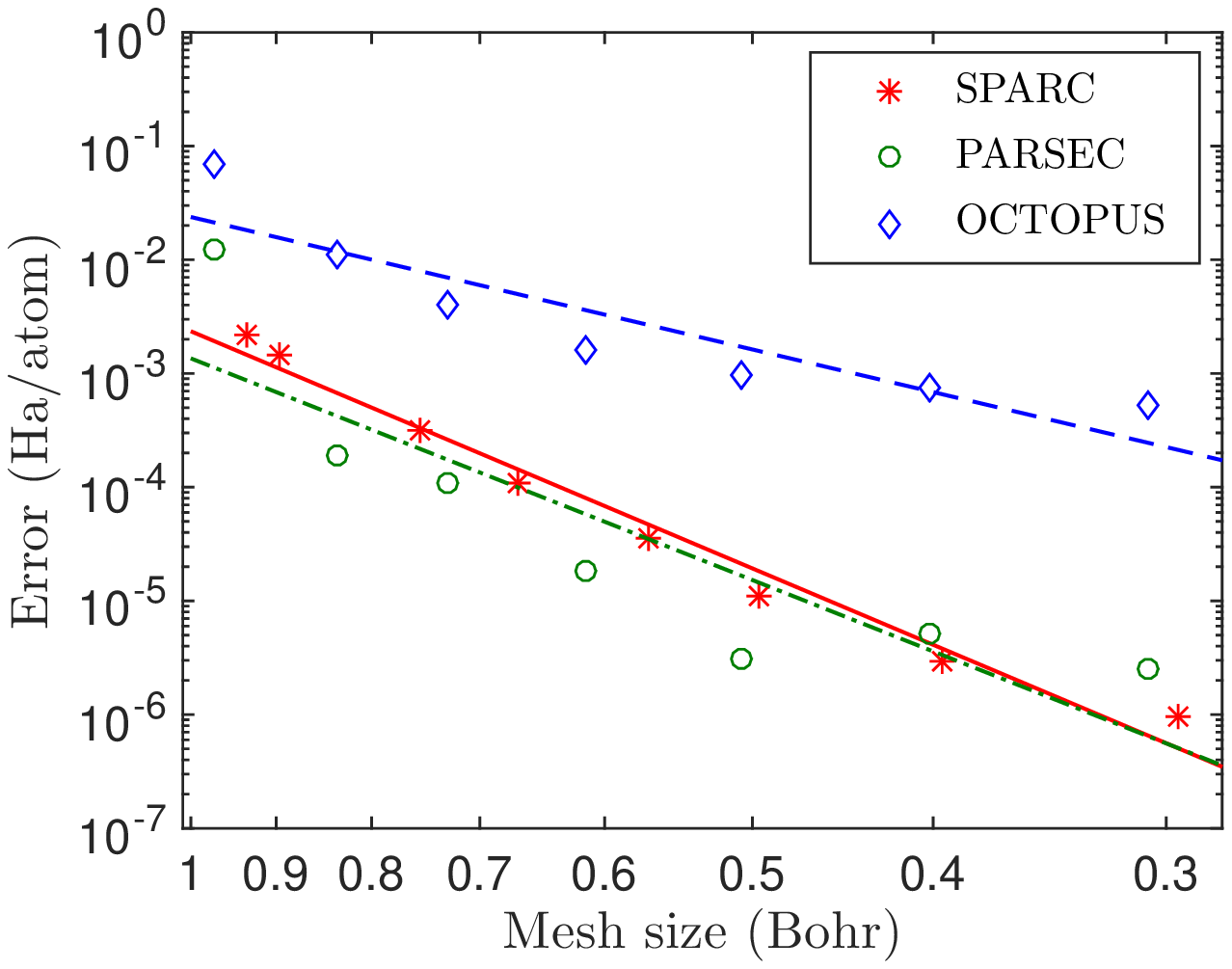}}
\subfloat[Forces]{\label{fig:forceConvergence_rs}\includegraphics[keepaspectratio=true,width=0.46\textwidth]{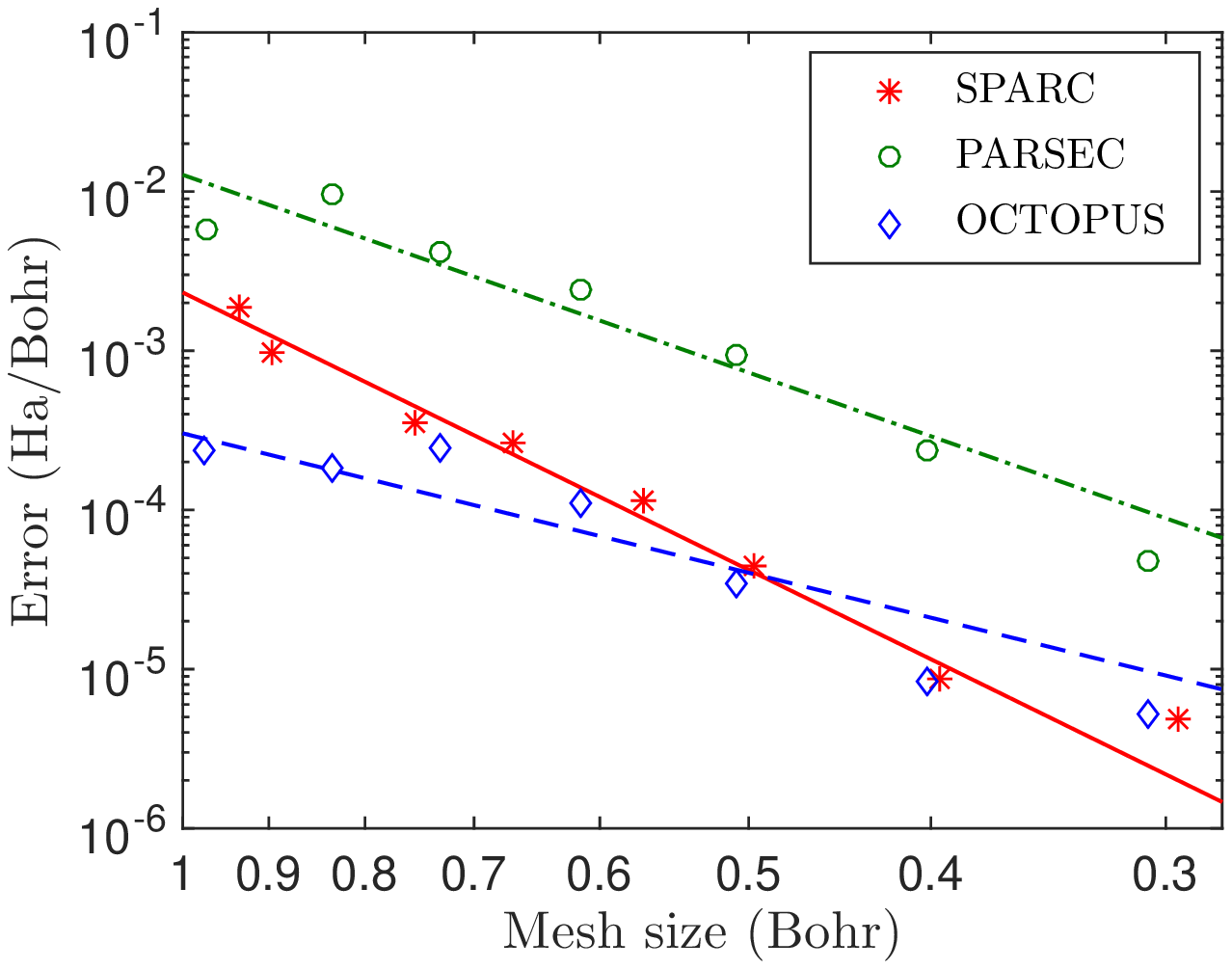}}
\caption{Convergence of the energy and atomic forces with respect to mesh size to reference planewave result for $3\times 3 \times 3$ FCC unit cells of aluminum with a vacancy. The error in energy for PARSEC and OCTOPUS has been defined with respect to the results obtained for a mesh-size of $h=0.201$ Bohr (See footnote \ref{footnote:OCT:PAR})}.
\label{fig:convergenceDiscretization_rs}
\end{figure}

We now study the accuracy and efficiency of SPARC relative to PARSEC \cite{chelikowsky1994finite} and OCTOPUS \cite{OCTOPUS}---two well-established DFT codes that employ the finite-difference discretization. First, we plot in Fig. \ref{fig:convergenceDiscretization_rs} the convergence of the energy and atomic forces as a function of mesh-size for the system consisting of $3\times 3 \times 3$ FCC unit cells of aluminum with a vacancy. The error has been defined with respect to ABINIT,\footnote{For crystals, the energy computed by PARSEC and OCTOPUS is found to converge to an answer that is significantly different from ABINIT. Therefore, the error in energy for PARSEC and OCTOPUS has been defined with respect to the results obtained for a mesh-size of $h=0.201$ Bohr. \label{footnote:OCT:PAR}} wherein we employ a plane-wave cutoff of $32$ Ha, which results in energy and forces that are converged to within $1 \times 10^{-6}$ Ha/atom and $1 \times 10^{-6}$ Ha/Bohr, respectively. We observe that the prefactors associated with the convergence of the energy in OCTOPUS and the forces in PARSEC are significantly larger than those for the other two codes. Notably, SPARC demonstrates the best convergence in terms of energy and forces amongst the three finite-difference packages. 

\begin{figure}[H]
\centering
\subfloat[Strong scaling]{\label{Fig:StrongScalingRS}\includegraphics[keepaspectratio=true,width=0.48\textwidth]{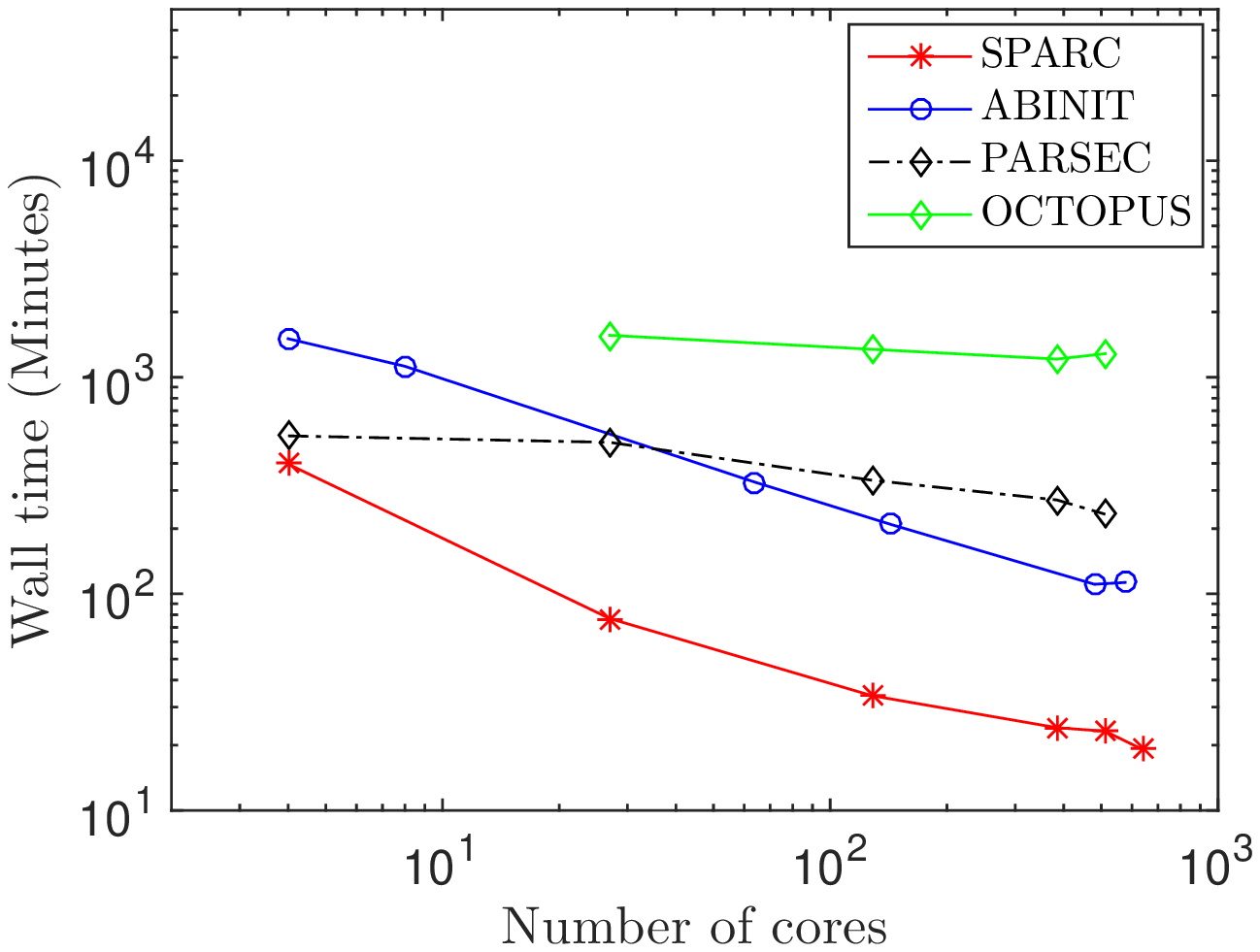}}
\subfloat[Weak scaling]{\label{Fig:WeakScalingRS}\includegraphics[keepaspectratio=true,width=0.48\textwidth]{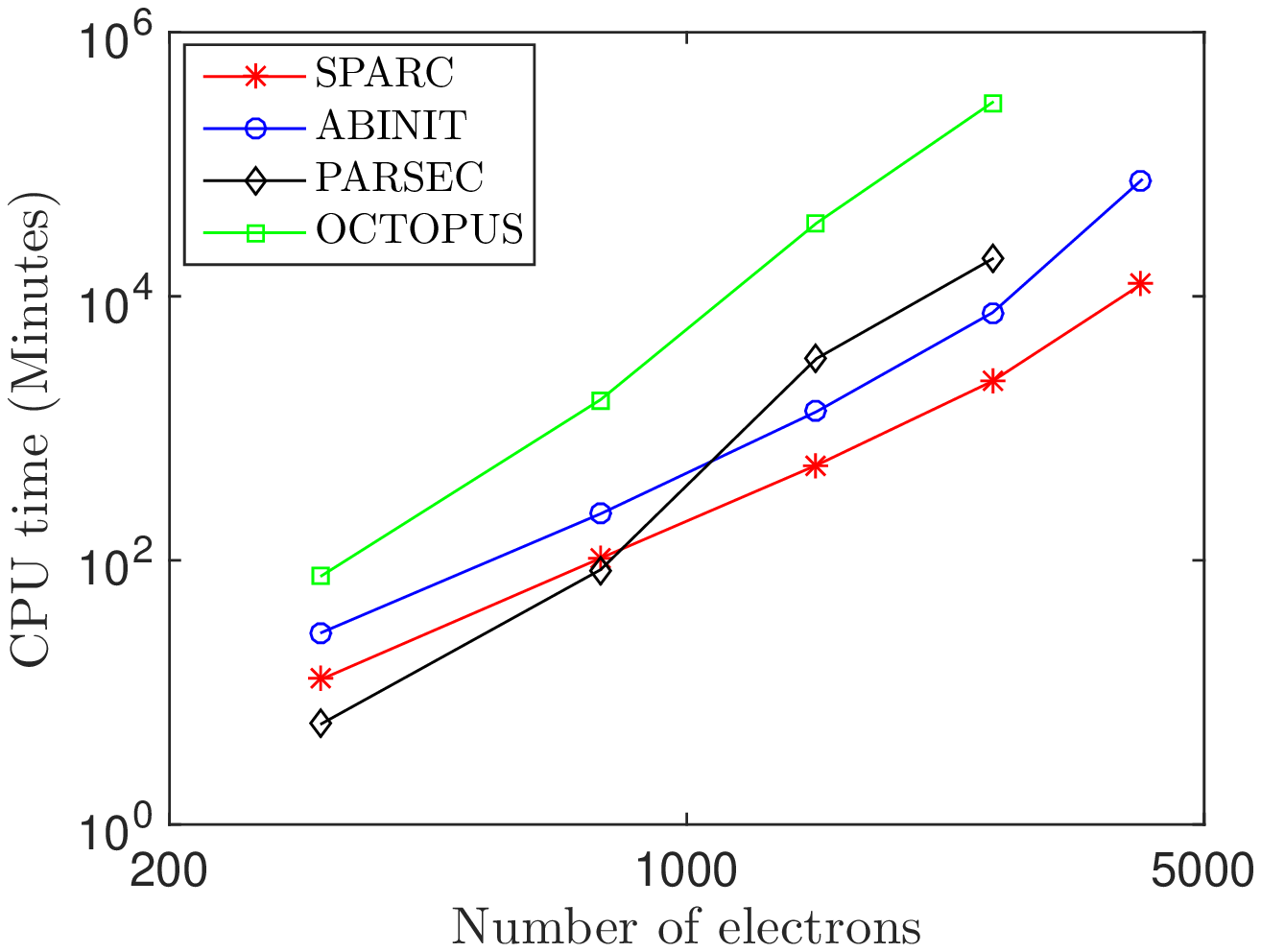}}
\caption{Strong and weak scaling for SPARC, PARSEC, OCTOPUS, and ABINIT. The system utilized for strong scaling is $6\times 6\times 6$ FCC unit cells of aluminum with a vacancy. The systems employed for weak scaling are $3\times 3\times 3$, $4\times 4\times 4$, $5\times 5\times 5$, $6\times 6\times 6$, and $7\times 7\times 7$ unit cells of aluminum, each with a vacancy.}
\label{Fig:Scaling:RealSpace}
\end{figure}

Next, we study the weak and strong scaling of SPARC relative to PARSEC and OCTOPUS. Specifically, we perform the weak and strong scaling tests described in Section \ref{Subsec:ScalingAndPerformance}, with a mesh of $h$ = $0.778$ Bohr employed in all three codes. We present the results obtained in Fig. \ref{Fig:Scaling:RealSpace}, wherein the time taken for the first SCF iteration has been excluded \cite{sparc2016cluster}. In the strong scaling test, the minimum wall time achieved by SPARC is smaller by factors of $12.2$, $93$, and $5.8$ relative to PARSEC, OCTOPUS, and ABINIT, respectively.\footnote{It is worth noting that the local electrostatic reformulation in SPARC significantly outperforms conventional Fourier based approaches for moderate to large systems, especially in the context of parallel computing. For example, in the strong scaling test, the time taken for the electrostatics in SPARC is a factor of $2.7$ smaller compared to OCTOPUS. This speedup increases to $62.6$ for the calculation on $384$ cores.} In the weak scaling test, the increase in CPU time with number of electrons for SPARC, PARSEC, OCTOPUS, and ABINIT is $\mathcal{O}(N_e^{2.64})$, $\mathcal{O}(N_e^{3.90})$, $\mathcal{O}(N_e^{3.98})$ and $\mathcal{O}(N_e^{2.99})$ respectively. It is clear that SPARC is able to outperform PARSEC because of significantly higher efficiency in strong scaling, and is able to outperform OCTOPUS because of significantly smaller prefactor and superior strong scaling.

Finally, we compare the minimum wall time---excluding the time for the first SCF iteration---that can be achieved by SPARC, PARSEC, OCTOPUS, and ABINIT for the aforementioned systems. From the results presented in Table \ref{Table:realspaceTime}, we find that SPARC demonstrates speedup by up to factors of 12.2, 72.3, and 5.55 compared to PARSEC, OCTOPUS, and ABINIT, respectively. Overall, these results demonstrate that SPARC is an efficient DFT formulation and implementation that is not only highly competitive with well-established plane-wave codes, but also significantly outperforms well-established finite-difference codes. 

\begin{table}[H]
\centering
\begin{tabular}{ccccc}
\hline
System &   SPARC     & PARSEC & OCTOPUS & ABINIT  \\  
       & $h=0.778$ Bohr & $h=0.778$ Bohr & $h=0.778$ Bohr & $E_{cut} = 9$ Ha \\
\hline
$3 \times 3 \times 3$        &   $0.36$ $(64)$  & $1.58$ $(64)$ & $3.64$ $(8)$ & $1.59$ $(180)$      \\
$4 \times 4 \times 4$       &   $1.47$ $(96)$  & $6.88$ $(128)$ & $25.33$ $(64)$ & $5.27$ $(320)$     \\
$5 \times 5 \times 5$     &   $6.39$ $(256)$  & $43.8$ $(256)$ & $465.24$ $(256)$ & $16.51$ $(396)$     \\
$6 \times 6 \times 6$     &   $19.06$ $(640)$ & $233$ $(512)$ & $1212.75$ $(384)$  & $108.77$ $(480)$  \\
$7 \times 7 \times 7$     &   $71.80$ $(704)$ & $705$ $(512)$ & $-$ & $388.67$ $(795)$  \\
\hline
\end{tabular}
\caption{Minimum wall time in minutes for $n\times n \times n$ ($n \in \{3,4,5,6,7 \}$) FCC unit cells of aluminum with a vacancy. The number in brackets represents the number of cores on which the minimum wall time is achieved.}
\label{Table:realspaceTime}
\end{table}


\end{document}